\documentclass[11pt,a4paper]{article}
\usepackage{fullpage,amssymb,amsmath,amsfonts,amsthm,paralist,hyperref,graphicx,color,enumitem,wrapfig,rotating}
\usepackage[normalem]{ulem}
\usepackage[vlined,ruled,longend]{algorithm2e}

\usepackage{float}
\makeatletter
\newcommand\floatc@mybox[2]{\vbox{\hbadness10000
\moveleft3.4pt\vbox{\advance\hsize by6.8pt
\hrule \hbox to\hsize{\vrule\kern3pt
\vbox{\kern3pt\vbox{\advance\hsize by-6.8pt{\@fs@cfont #1} #2}\kern3pt}\kern3pt\vrule}}}}%
\newcommand\fs@mybox{\def\@fs@cfont{\bfseries}\let\@fs@capt\floatc@mybox
\def\@fs@pre{\setbox\@currbox\vbox{\hbadness10000
\moveleft3.4pt\vbox{\advance\hsize by6.8pt
\hrule \hbox to\hsize{\vrule\kern3pt
\vbox{\kern4.5pt\box\@currbox\kern4.5pt}\kern3pt\vrule}\hrule}}}%
\def\@fs@mid{}%
\def\@fs@post{}%
\let\@fs@iftopcapt\iftrue}
\makeatother
\floatstyle{mybox}
\newfloat{inset}{tb}{ins}
\floatname{inset}{Inset}

\newcommand{\Reals}{\mathbb{R}}
\def\smallminus{\hbox{\footnotesize-}}
\newcommand\descr[1]{{\ensuremath{\thinmuskip=0mu\let\m\smallminus#1}}}
\def\fwd[#1]{\ensuremath{\left[\begin{smallmatrix}#1\end{smallmatrix}\right\}}}
\def\rev[#1]{\ensuremath{\left\{\begin{smallmatrix}#1\end{smallmatrix}\right]}}
\def\edge[#1]{\ensuremath{\begin{smallmatrix}#1\end{smallmatrix}}}
\def\vtx#1{\ensuremath{\left(\begin{smallmatrix}#1\end{smallmatrix}\right)}}

\newcommand{\XSays}[3]{{\color{#2}
      {$\rule[-0.12cm]{0.2in}{0.5cm}$\fbox{\tt
            #1:} }%
      \itshape #3
      \marginpar{\color{#2}\tt #1}%
      \def\comment{#3}\def\empty{}\ifx\comment\empty\else
      {$\rule[0.1cm]{0.3in}{0.1cm}$\fbox{\tt
            end}$\rule[0.1cm]{0.3in}{0.1cm}$} \fi
   }%
}

\title{Sixteen space-filling curves and traversals\\for $d$-dimensional cubes and simplices}

\author{Herman Haverkort\thanks{Dept. of Mathematics and Computer Science, Eindhoven University of Technology, the Netherlands,\hfill\break cs.herman@haverkort.net}}

\begin{document}
\maketitle

\begin{abstract}
This article describes sixteen different ways to traverse $d$-dimensional space recursively in a way that is well-defined for any number of dimensions. Each of these traversals has distinct properties that may be beneficial for certain applications.
Some of the traversals are novel, some have been known in principle but had not been described adequately for any number of dimensions, some of the traversals have been known. This article is the first to present them all in a consistent notation system. Furthermore, with this article, tools are provided to enumerate points in a regular grid in the order in which they are visited by each traversal.
In particular, we cover:\begin{itemize}
\item five discontinuous traversals based on subdividing cubes into $2^d$ subcubes: Z-traversal (Morton indexing), U-traversal, Gray-code traversal, Double-Gray-code traversal, and Inside-out traversal;
\item two discontinuous traversals based on subdividing simplices into $2^d$ subsimplices: the Hill-Z traversal and the Maehara-reflected traversal;
\item five continuous traversals based on subdividing cubes into $2^d$ subcubes: the Base-camp Hilbert curve, the Harmonious Hilbert curve, the Alfa Hilbert curve, the Beta Hilbert curve, and the Butz-Hilbert curve;
\item four continuous traversals based on subdividing cubes into $3^d$ subcubes: the Peano curve, the Coil curve, the Half-coil curve, and the Meurthe curve.
\end{itemize}
All of these traversals are self-similar in the sense that the traversal in each of the subcubes or subsimplices of a cube or simplex, on any level of recursive subdivision, can be obtained by scaling, translating, rotating, reflecting and/or reversing the traversal of the complete unit cube or simplex.
\end{abstract}

\section{Introduction}
\begin{figure}[t]
\centering
\includegraphics[width=0.9\hsize]{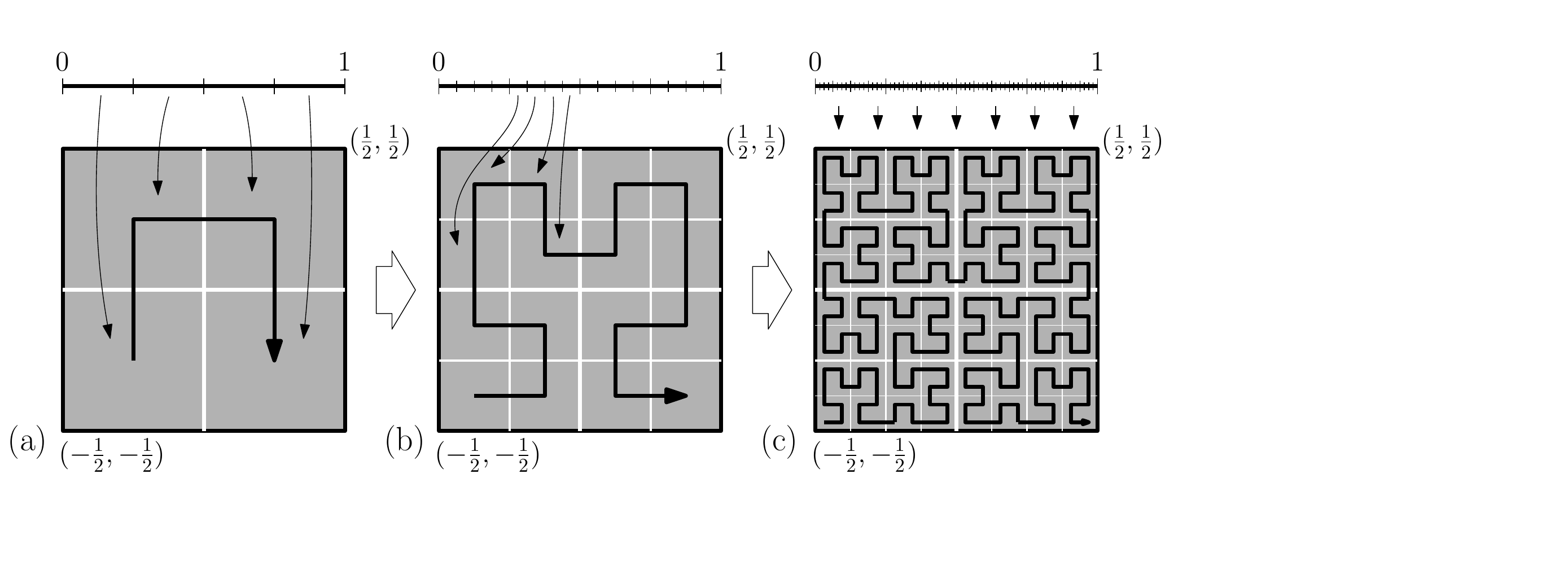}
\caption{A sketch of Hilbert's space-filling curve.}
\label{fig:hilbert2d}
\end{figure}
A space-filling curve in $d$ dimensions is a continuous, surjective mapping from $\Reals$ to $\Reals^d$.
One of my favourite examples is Hilbert's curve~\cite{Hil1891}, because it demonstrates the core ideas underlying many space-filling curves. Hilbert's curve can be described as a recursive construction that maps the unit interval $[0,1]$ to the unit square; for ease of notation, in this paper, we work with a unit square $[-\frac12,\frac12]^2$ centred on the origin. The unit square is divided into a grid of $2 \times 2$ square cells, while the unit interval is subdivided into four subintervals. Each subinterval is then matched to a cell; thus Hilbert's curve traverses the cells one by one in a particular order. The procedure is applied recursively to each subinterval-cell pair, so that within each cell, the curve makes a similar traversal (see Figure~\ref{fig:hilbert2d}). By carefully reflecting and/or rotating the traversals within the cells, one can ensure that each cell's first subcell touches the previous cell's last subcell.

The result is a fully-specified, continuous, surjective mapping $\tau$ from the unit interval to the unit square. The mapping is \emph{measure-preserving}: for any set of points $S \subset [0,1]$ with one-dimensional Lebesgue measure $z$, the image $\bigcup_{x\in S} \tau(x)$ of $S$ has $d$-dimensional Lebesgue measure~$z$. Moreover, the mapping usually \emph{preserves locality}: points that are close to each other in $d$-dimensional space usually lie close to each other along the curve, and vice versa.

Since the 19th century, various space-filling curves have appeared in the literature~\cite{HW10,Sag94}, along with discontinuous measure-preserving mappings~\cite{Sam06}. We will collectively call these mappings \emph{space-filling traversals}. Different traversals have different properties that make them suitable for different applications to indexing of multidimensional points~\cite{ARRWW97,KF93,LK01,LLL01,Sam06}, geometric algorithms and data structures~\cite{Buc09,HT16}, finite element methods~\cite{Bad13}, load balancing in parallel computing~\cite{BMW06,HKRSW12}, improving cache utilization in computations on large matrices~\cite{BZ06} or in image rendering~\cite{Voo91}, combinatorial optimization~\cite{PB89}, image compression~\cite{AM90}, information visualization~\cite{AHLRSS13}, sonification~\cite{Gro07,Vog10}, and musical composition~\cite{Hav17a}---to give only a few examples of applications and references. The function of the space-filling traversal in such applications is typically to provide an order in which to process points in a two- or higher-dimensional space.

However, in many cases, space-filling traversals have been described only for two- or sometimes three-dimensional spaces. It is often clear that the traversals can be generalized to higher dimensions in one way or another, but it is often not clear exactly which rotations and reflections are needed to maintain continuity or other useful properties of the mapping. When adequate descriptions exist, they are found in various original sources that ease use their own notation system, which can make it hard to compare the traversals to each other and to implement them correctly in software. Figures in two or three dimensions do not usually suffice to define a natural and unambiguous generalization to higher dimensions.

In this work, I provide precise descriptions of how various space-filling traversals could be generalized to higher dimensions. For this purpose I use the notation system, originally based on ideas by Arie Bos, that I introduced in my recent work on three-dimensional Hilbert curves~\cite{Hav17}.

This article is structured as follows. Section~\ref{sec:notation} describes the notation system
and discusses a basic pattern that is common to many of the traversals in this paper. Next are the traversals themselves: discontinuous traversals for cubes~(Section~\ref{sec:discontinuous}) and simplices (Section~\ref{sec:simplices}), continuous traversals based on subdivision into $2^d$ cubes (Section~\ref{sec:Hilbert}), and continuous traversals based on subdivision into $3^d$ cubes (Section~\ref{sec:Peano}). Section~\ref{sec:lowd} discusses several traversals that have, so far, evaded generalization to more than two or three dimensions. Section~\ref{sec:squared} explains how higher-dimensional traversals can be constructed by \emph{squaring} (repeated application) of lower-dimensional traversals. Section~\ref{sec:practical} briefly discusses basic implementation issues and a software prototype to generate the traversals. Finally, we discuss unanswered questions in Section~\ref{sec:conclusions}.

\section{Defining self-similar traversals}\label{sec:notation}
This section describes the notation that is used in this paper to define self-similar traversals. Readers who are familiar with my previous work will find that much of this section is almost the same as in my article on three-dimensional Hilbert curves~\cite{Hav17}. Compared to my previous work, the main adaptations are: (i) updated definitions to cover traversals based on subdivision in any number of subcubes; (ii) a less rigorous approach to figures (for higher dimensions, they have limited use); (iii) additional notation to express functions of subcube coordinates.

\paragraph{Mapping the unit interval to the unit cube}\label{sec:mapping}
We can define a \emph{self-similar traversal} of points in a $d$-dimensional cube as follows. We consider the unit cube $C$ to be subdivided into $D = s^d$ subcubes of equal size, for some integer \emph{scale factor}~$s$. We specify a \emph{base pattern}: an order in which the traversal visits these subcubes. Let $C_1,...,C_D$ be the subcubes indexed by the order in which they are visited. Moreover, we specify, for each subcube $C_i$, a transformation~$\sigma_i$ that maps the traversal of the cube as a whole to the traversal of $C_i$. Each $\sigma_i$ can be thought of as a triple $(\gamma_i, \rho_i, \chi_i)$, where $\gamma_i: C \rightarrow C$ is one of the $2^d d!$ symmetries of the unit cube, $\rho_i: C \rightarrow C_i$ translates the unit cube and scales it down to map it to $C_i$, and $\chi_i: [0,1] \rightarrow [0,1]$ is a function that specifies whether or not to reverse the direction of the traversal: it is defined by $\chi_i(t) = t$ for a forward traversal, and by $\chi_i(t) = 1-t$ for a reversed traversal.

As illustrated in Figure~\ref{fig:hilbert2d}, we can think of a traversal as mapping segments of the unit interval to subcubes of the unit cube~$C$. For a given level of refinement $k$, consider the unit interval subdivided into $D^k$ segments of equal length, and the unit cube subdivided into $D^k$ subcubes of equal size. Let $s(i,k)$ be the $i$-th segment of the unit interval, that is, the interval $[(i-1)/D^k, i/D^k]$. Let $c(i,k)$ be the $i$-th subcube in the traversal. We can determine $c(i,k)$ from the transformations $\gamma, \rho$ and $\chi$ as follows. If $k = 0$, then $i$ must be 1 and $c(i,k) = C$. Otherwise, let $z = D^{k-1}$ be the number of subcubes within a first-level subcube, let $b = \lceil i/z\rceil$ be the index of the first-level subcube that contains $c(i,k)$, and let $j$ be the index of $c(i,k)$ within~$C_b$. More precisely, if $\chi_b$ indicates a forward traversal of $C_b$, then $j = i - (b-1)z$, and if $\chi_b$ indicates a reverse traversal of $C_b$ then $j = bz - i + 1$. Then we have $c(i,k) = \rho_b(\gamma_b(c(j,k-1)))$, and the traversal maps the segment $s(i,k)$ to the cube~$c(i,k)$.

As $k$ goes to infinity, the segments $s(i,k)$ and the cubes $c(i,k)$ shrink to points, and the traversal defines a mapping from points on the unit interval to points in the unit cube. By construction, the mapping is surjective. However, it may be ambiguous, as some points in the unit interval lie on the boundary between segments for any large enough $k$. We may break the ambiguity towards the left or towards the right, by considering segments to be relatively open on the left or on the right side, respectively. In the first case, for a given $k$, we consider a point $t$ on the unit interval to be part of the $i$-th interval with $i = \lceil D^k t\rceil$, and we define a mapping $\tau^{-}: (0,1] \rightarrow C$ to points in the unit cube by $\tau^{-}(t) = \lim_{k\rightarrow\infty} c(\lceil D^k t\rceil, k)$. In the second case, we consider $t$ to be part of the $i$-th interval with $i = \lfloor D^k t\rfloor + 1$, and we define a mapping $\tau^{+}: [0,1) \rightarrow C$ by $\tau^{+}(t) = \lim_{k\rightarrow\infty} c(\lfloor D^k t\rfloor + 1, k)$.

\paragraph{Space-filling curves}\label{sec:selfsimilarcurves}
Suppose a traversal has the property that consecutive segments of the unit interval are always matched to subcubes that touch each other. Then, as $k$ increases, the up to two subcubes corresponding to the segments that share a point $t \in [0,1]$ must shrink to the same point $p \in C$. For $t \in (0,1)$, we thus have $\tau^{-}(t) = \tau^{+}(t)$. Moreover, the functions $\tau^{-}$ and $\tau^{+}$ are continuous. The traversal thus follows a \emph{space-filling curve}. We say a space-filling curve is \emph{face-continuous} if, for any $0 \leq a < b \leq 1$, the interior of $\bigcup_{a < t < b} \tau^+(t)$ is connected. We say a traversal is \emph{semi-face-continuous} if the interior of any section $\bigcup_{a < t < b} \tau^+(t)$ consists of at most a constant number of connected components.

Note that the traversals discussed in this article are measure-preserving functions from the unit interval to a subset of $\Reals^d$. We will not discuss traversals that cannot be refined recursively. For example, traversals of grid points in lexicographical order of their coordinates (row-major order) or in order of distance from the origin are not among those treated in this article.

\paragraph{Defining self-similar traversals by signed permutations}\label{sec:definitionbypermutations}
To define the mappings $\tau^{-}$ and $\tau^{+}$, all we need to do is to specify, for each $i \in \{1,...,D\}$, the transformation $\gamma_i$, the location of $C_i$ (or, to the same effect, $\rho_i$), and the orientation function $\chi_i$.

\begin{figure}
\centering
\includegraphics{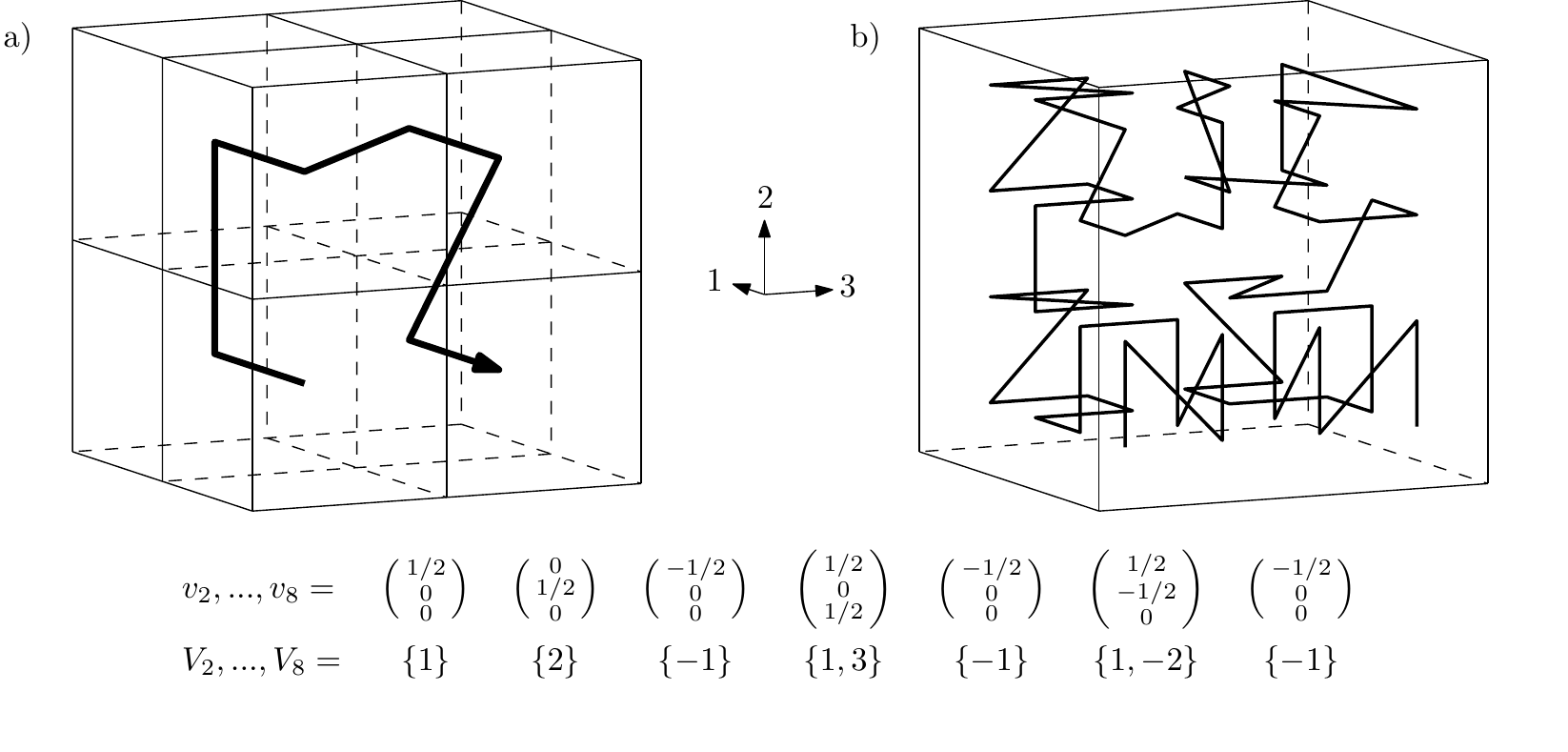}\vskip-0.5\baselineskip
\caption{(a) Example of a base pattern, sketched by a curve that connects the subcube centres in order, with the corresponding vectors $v_2,...,v_8$ and ``moves'' $V_2,...,V_8$. Note how, for positive $j$, a set $V_i = \{ j \}$ can be interpreted as: move forward along the $j$-th coordinate axis to get from $C_{i-1}$ to $C_i$, while $V_i = \{ -j \}$ means: move back along the $j$-th coordinate axis, and $V_i = \{ j_1, j_2 \}$ indicates a diagonal move, simultaneously moving in coordinates $|j_1|$ and $|j_2|$. (b) Example of the order in which a curve with this base pattern might visit the centre points of the subcubes on the second level of subdivision. The traversals within the first-level subcubes can be described by the following  permutations, in order from $C_1$ to $C_8$:
$\{2,3,-1],
[3,1,2\},
[3,1,2\},
\{-1,-2,-3],
\{1,-2,-3],
\{-1,-3,2],
\{-3,1,2],
[2,-3,-1\}$.}
\label{fig:notation}
\end{figure}

We specify the base pattern by indicating, for each subcube $C_i$ with $1 < i \leq D$, where $C_i$ lies relative to the previous subcube $C_{i-1}$. Let $c_i$ be the centre point of $C_i$ (relative to the origin of the unit cube); the position of $C_i$ relative to $C_{i-1}$ can then be expressed by the vector $v_i = c_i - c_{i-1}$. We use square brackets to index the elements of a vector, so $v_i$ is a column vector with elements $v_i[1], v_i[2], ... v_i[d]$. However, in our notation system, we specify $v_i$ in a more compact way, namely by a (multi-)set $V_i$ of numbers from $\{-1,...,-d\} \cup \{1,...,d\}$. Each number $e \in V_i$ represents a step parallel to coordinate axis $|e|$ in the direction specified by the sign of $e$. More precisely, we define the $d$-dimensional vector $u_e$ by $u_e[j] = 0$ for all $j \neq |e|$, and $u_e[|e|] = \mathrm{sign}(e)/s$ (recall that $1/s$ is the width of a subcube). The set $V_i$ now defines $v_i$ as follows: $v_i = \sum_{e \in V_i} u_e$. For an example, see Figure~\ref{fig:notation}(a).

Assume the unit cube is centred at the origin. Each transformation $\gamma_i: C \rightarrow C$ is a symmetry of the unit cube and can be interpreted as a matrix $M_i$ such that $\gamma_i(x) = M_i x$, where $x$ is a point given as a column vector of its coordinates. Each row and each column of~$M_i$ contains exactly one non-zero entry, which is either $1$ or $-1$. We specify such a matrix by a signed permutation of row indices, that is, a sequence of numbers $\Pi_i = \pi_i[1],...,\pi_i[d]$ whose absolute values are a permutation of $\{1,...,d\}$, and that corresponds to the matrix in the following way: the non-zero entry of column $j$ is in row $|\pi_i[j]|$ and has the sign of~$\pi_i[j]$. We write the sequence $\pi_i[1],...,\pi_i[d]$ between $[$ and $\}$ to specify a forward traversal ($\chi_i(t) = t$), whereas we write the sequence $\pi_i[1],...,\pi_i[d]$ between $\{$ and $]$ to specify a reverse traversal ($\chi_i(t) = 1-t$). For an example, see Figure~\ref{fig:notation}(b).

A complete self-similar traversal is now specified by listing the signed and directed permutations $\Pi_1,...,\Pi_D$, with, between each pair of consecutive permutations $\Pi_{i-1}$ and $\Pi_i$, the set $V_i$ that gives the location of $C_i$ relative to $C_{i-1}$. Depending on lay-out requirements, we may omit commas and/or we may write the numbers of a set $V_i$ or a signed permutation $\Pi_i$ below each other instead of from left to right; we also omit braces around $V_i$. Thus we get the following description of the traversal from Figure~\ref{fig:notation}(b):\[\descr{%
\rev[2\\3\\\m1]\edge[1]
\fwd[3\\1\\2]\edge[2]
\fwd[3\\1\\2]\edge[\m1]
\rev[\m1\\\m2\\\m3]\edge[1\\3]
\rev[1\\\m2\\\m3]\edge[\m1]
\rev[\m1\\\m3\\2]\edge[1\\\m2]
\rev[\m3\\1\\2]\edge[\m1]
\fwd[2\\\m3\\\m1]
}\]
Note that we do not specify the location of $C_1$ explicitly, but it can be derived from the sets $V_2,...,V_D$: we place the pattern such that the mean of the subcubes' centre points is the centre of the unit cube.

\paragraph{Expressing permutations as a function of subcube coordinates}
To be able to define traversals for any number of dimensions, we have to define the permutations in another way than by writing them out explicitly. Therefore, in the following sections, a permutation $\Pi_i$ will typically be expressed as a function of $i$ and/or $c_i$, the centre of the subcube to which it is applied. For this purpose we will use the notation $\mathring{c}_i[j]$ to denote the sign of $c_i[j]$.

\paragraph{Well-folded traversals}\label{sec:graycode}
Most of the traversals discussed in this paper actually have the same base pattern, based on the \emph{binary reflected Gray code}~\cite{Gry53} (the differences between the traversals are in the transformations applied within the subcubes). The base pattern of four other traversals is based on the \emph{ternary reflected Gray code}. These patterns can be described as follows. Let $\{e\}$ be a set representing a vector $u_e$ that indicates a move from one subcube centre to the next, as described above. Given a sequence of such sets $E = \{e_1\}, ..., \{e_k\}$, its reverse $\overleftarrow{E}$ is obtained by reversing the order of the sets and changing the signs, that is, $\overleftarrow{E} = \{-e_k\}, ..., \{-e_1\}$. Now let $G_2(0)$ and $G_3(0)$ be an empty sequence. We define $G_2(k)$, for $k = \{1,2,3,...,d\}$, as the concatenation of $G_2(k-1)$, $\{k\}$, and $\overleftarrow{G_2(k-1)}$; we define $G_3(k)$ as the concatenation of $G_3(k-1)$, $\{k\}$, $\overleftarrow{G_3(k-1)}$, $\{k\}$, and $G_3(k-1)$ once more. The base pattern of most traversals starts in the subcube that lies on the low side with respect to all coordinates, and then follows the sequence of moves $G_2(d)$ or $G_3(d)$. For examples, see all traversals in Sections \ref{sec:discontinuous}, \ref{sec:Hilbert} and~\ref{sec:Peano}, except the Z-traversal.

Bos and I call traversals with base pattern $G_2(d)$ \emph{well-folded}~\cite{BH16}. In that case all coordinates of subcube centres have one of only two values, $-\frac14$ and $\frac14$. Therefore, a subcube centre $c$ can be interpreted as a binary number $\mathrm{bin}(c)$ with digits $b[d]...b[1]$, where $b[j] = 0$ if $c[j] < 0$ and $b[j] = 1$ if $c[j] > 0$. Let $g(n)$ be a function that returns the result of applying the bitwise exclusive-or operation to the binary representations of $n$ and $\lfloor n/2\rfloor$. One can now show that, for well-folded curves, $\mathrm{bin}(c_i) = g(i-1)$ (note that $i$ runs from $1$ to $2^d$ in this exposition, hence the need to substract 1 in the argument to $g$). Conversely, for a subcube that contains a unit cube vertex with coordinate vector $c$, we can obtain its rank in the traversal order as $g^{-1}(\mathrm{bin}(c)) + 1$, where $g^{-1}$ is the inverse of $g$.

\begin{figure}
\raggedleft
\includegraphics[width=\hsize]{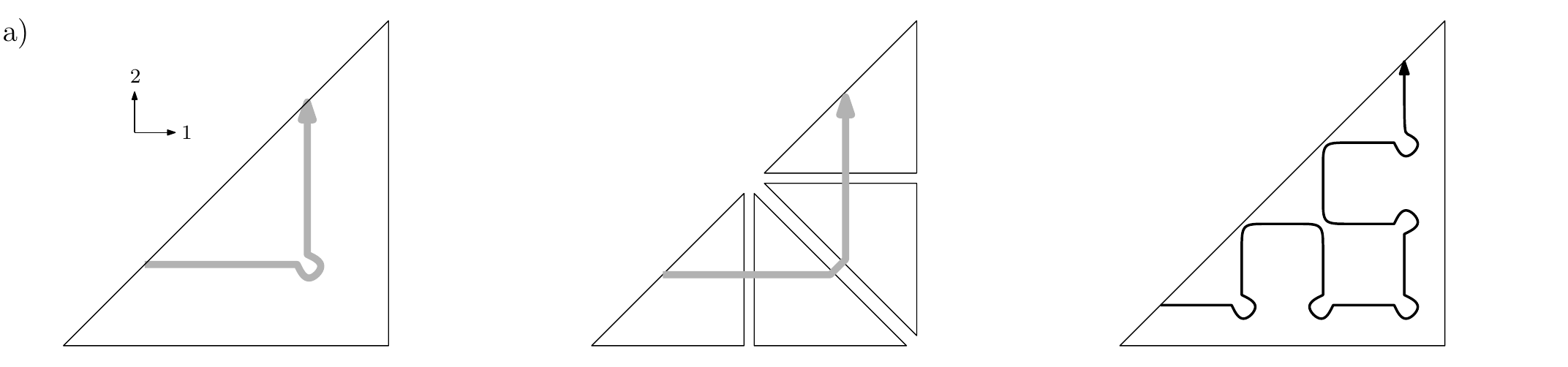}\\
$\descr{\fwd[1\\2]\edge[1]\rev[\m1\\2]\fwd[2\\\m1]\edge[2]\rev[\m2\\\m1]}$
\par\addvspace\baselineskip
\includegraphics[width=\hsize]{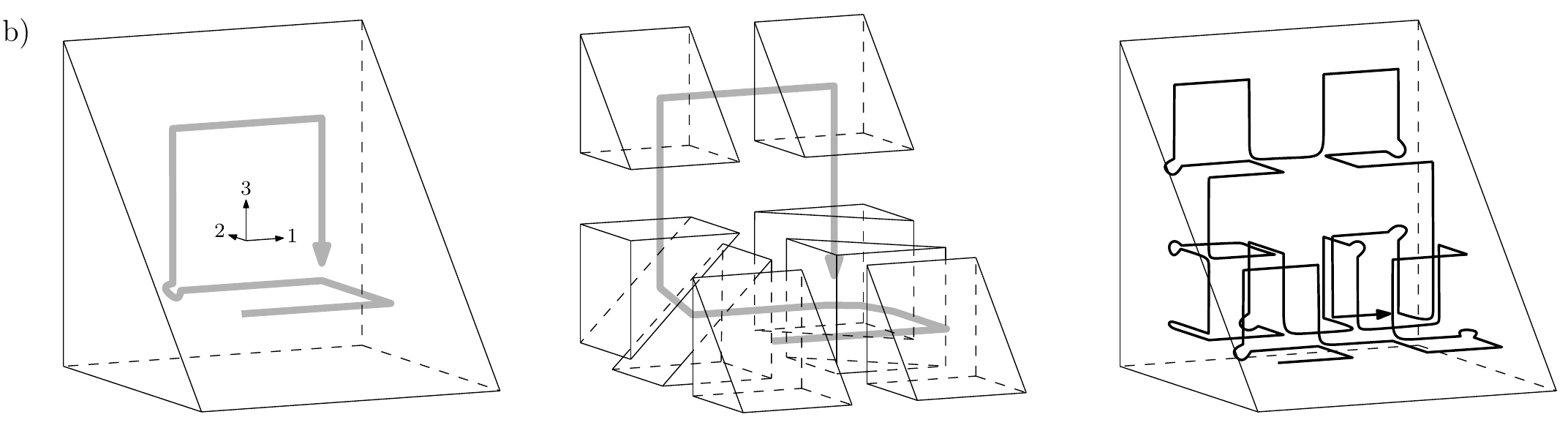}\\
$\descr{\fwd[1\\2\\3]\edge[1]\rev[\m1\\\m3\\\m2]\edge[2]\fwd[\m3\\\m1\\2]\edge[\m1]\rev[\m2\\1\\3]\fwd[\m2\\3\\1]\edge[3]\fwd[1\\2\\3]\edge[1]\rev[\m1\\2\\3]\edge[\m3]\fwd[\m3\\2\\\m1]}$
\caption{Two curves that fill non-cubic shapes. For both curves, the figure shows how the shape is subdivided into tiles; it shows a curve that sketches the order in which these tiles are visited; on the right, a curve that sketches the order in which the centres of the tiles in the second level of subdivision are visited; and, finally, the definition of the traversal in our notation based on signed permutations. Note that there are often multiple visits to the same centre points, since each visit corresponds to filling only half of the corresponding square or cube. In the curve definitions, staying at a centre point for a second visit is indicated by an empty move between two signed permutations.
(a) The P\'olya curve~\cite{Pol1913} that fills an isosceles right triangle (equivalent to half of Sierpi\'nski's curve~\cite{Sie75}).
(b) A curve that fills the extrusion of an isosceles right triangle~\cite{Hav17}.}
\label{fig:trianglecurves}
\end{figure}

\paragraph{Traversals of other shapes than cubes}
Our notation can also be used to describe traversals that are based on subdividing into other shapes than cubes, such as simplices or prisms, as long as these are still arranged in a regular cubic grid pattern. As the centre point for a simplex or prism, we take the centre of its bounding box. Figure~\ref{fig:trianglecurves} shows two examples.

\section{Discontinuous quadrant-by-quadrant traversals}\label{sec:discontinuous}

In this section we discuss five traversals that are based on subdividing cubes into $2^d$ subcubes. The traversals do not employ rotations in the subcubes (only translation, scaling and reflection), and the traversals are symmetric, so that reversals can be simulated by reflections. Therefore the traversals can be expressed not only with our notation system from Section~\ref{sec:notation}, but they can also be expressed (and implemented) relatively easily based on interleaving the bits of the binary representations of subcube coordinates.

More precisely, we identify any subcube (on any level of recursion) by the coordinates of the corner with the smallest coordinates. Thus all coordinates are at least~$-1/2$ and strictly less than~$1/2$. Consider each coordinate $x$ to be given as a string of digits of the fractional part of the binary representation of $x + 1/2$. For a given corner that represents a subcube, consider its $d$ coordinates given by a 0-1 matrix $X$ of $d$ rows, such that row $d+1-i$, that is, the $i$-th row from the bottom, contains the digits of the $i$-th coordinate, in order from most significant to least significant. For example, the subcube with coordinates $[-7/16,-6/16] \times [3/16,4/16] \times [-2/16,-1/16]$ would be represented by the coordinates $(0.0001, 0.1011, 0.0110)$, and thus:\[
  X = \left[\begin{smallmatrix}0&1&1&0\\1&0&1&1\\0&0&0&1\end{smallmatrix}\right].
\]
We may interpret the rows, columns, or the whole matrix $X$ as a single string of bits, so that we can talk about applying $g$ or $g^{-1}$ to it (see \emph{well-folded traversals} in Section~\ref{sec:notation}). In the case of applying operations to the whole matrix, we consider the string that contains the digits of $X$ column by column from left to right, and within each column, row by row from the top down.

As building blocks for traversals in this section we use the following operations on $X$:

\medskip\bgroup\def\arraystretch{1.75}
\begin{tabular}{ll}
\emph{inversion}: flip all bits in every second column & (applied to example $X$: $\left[\begin{smallmatrix}0&0&1&1\\1&1&1&0\\0&1&0&0\end{smallmatrix}\right]$) \\
\emph{row-coding}: apply $g$ to each row               & (applied to example $X$:
$\left[\begin{smallmatrix}0&1&0&1\\1&1&1&0\\0&0&0&1\end{smallmatrix}\right]$) \\
\emph{ranking}: apply $g^{-1}$ to the whole matrix $X$ & (applied to example $X$:
$\left[\begin{smallmatrix}0&0&1&0\\1&0&0&1\\1&0&0&0\end{smallmatrix}\right]$) \\
\emph{column-ranking}: apply $g^{-1}$ to each column   & (applied to example $X$:
$\left[\begin{smallmatrix}0&1&1&0\\1&1&0&1\\1&1&0&0\end{smallmatrix}\right]$) \\
\end{tabular}\egroup

\medskip
Each of the traversals in this section can be expressed as follows: for any given subcube corner, we perform a sequence of the aforementioned operations on its coordinate matrix $X$, and then we interpret the string with all digits of the matrix as a binary number that indicates the position of the subcube in the traversal. For example, if the sequence of operations consists of only column-ranking, the position index of our example subcube would be $011\,111\,100\,010$.

To obtain the order in which subcubes are visited, we sort their corners by these binary numbers. Note that initially, coordinates are put in $X$ row by row, but ultimately we interpret $X$ column by column, thus interleaving the bits (digits) of the coordinates. For each of the traversals discussed in this section, we specify the sequence of operations on $X$ that needs to be done to obtain the matrix whose column-by-column reading constitutes the correct subcube index.
\def\HowTo#1{ \textit{(Matrix implementation: #1.)}}

\begin{figure}
\centering\leavevmode
\includegraphics[width=0.3\hsize,page=7]{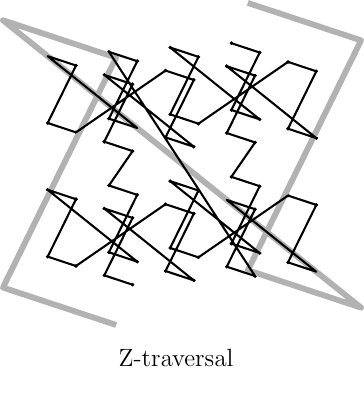}\hfill
\includegraphics[width=0.3\hsize,page=8]{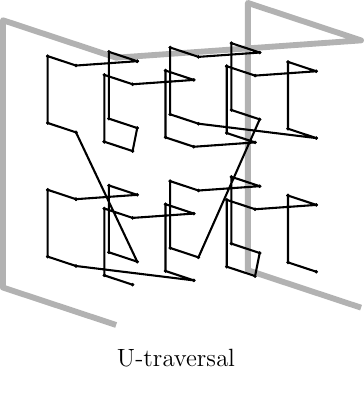}\hfill
\includegraphics[width=0.3\hsize,page=9]{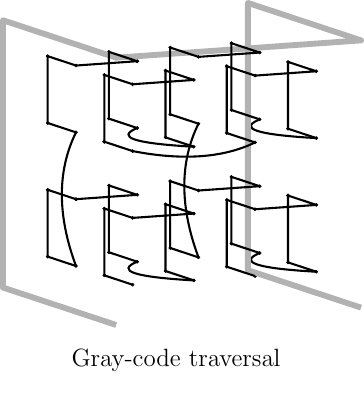}\\[\baselineskip]
\includegraphics[width=0.3\hsize,page=10]{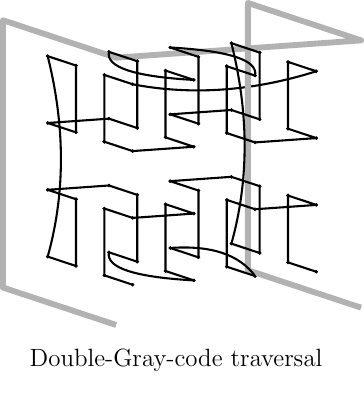}\quad
\raisebox{0.25\hsize}{\includegraphics[scale=1.3,page=19]{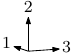}}\quad
\includegraphics[width=0.3\hsize,page=11]{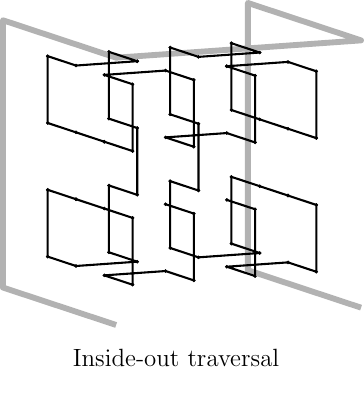}
\caption{The three-dimensional versions of the discontinuous traversals in this paper that are based on subdividing cubes into $2^d$ subcubes. The fat, gray curve indicates the order in which the vertices of the unit cube are visited; the thin, black curve indicates the order in which the centres of the subcubes in the second level of subdivision are visited.}\label{fig:discontinuous}
\end{figure}

\paragraph{Z-traversal}\HowTo{use as is}
The position of a subcube in the order is simply obtained by interleaving the coordinates. This results in a traversal with a zig-zag pattern, without any reflections. For example, with $d = 3$ we get (see Figure~\ref{fig:discontinuous}):\[
\descr{%
\fwd[1\\2\\3]\edge[1]%
\fwd[1\\2\\3]\edge[\m1\\2]%
\fwd[1\\2\\3]\edge[1]%
\fwd[1\\2\\3]\edge[\m1\\\m2\\3]%
\fwd[1\\2\\3]\edge[1]%
\fwd[1\\2\\3]\edge[\m1\\2]%
\fwd[1\\2\\3]\edge[1]%
\fwd[1\\2\\3]}.
\]
The Z-traversal is semi-face-continuous: the number of spatially connected components in any contiguous section of the curve is at most two, regardless of the number of dimensions~\cite{Bur15}. If subcube $B$ dominates subcube $A$ in the sense that all coordinates of $A$ are equal to those of $B$ or less, then $B$ is visited after $A$ (this property is not shared by any other traversal discussed in this paper).

The Z-traversal has long been known in computer science as Morton indexing~\cite{Mor66}. In fact, the subcubes are visited in the order in which they appear along Lebesgue's space-filling curve~\cite{Leb1904}. Lebesgue's curve would bridge the discontinuities in the Z-traversal by introducing connecting line segments, obtaining a mapping that is continuous but no longer measure-preserving.

\paragraph{U-traversal}\HowTo{column-ranking}
This traversal follows the well-folded base pattern in every subcube, without any reflections.
Example for $d = 3$ (see Figure~\ref{fig:discontinuous}):\[
\descr{%
\fwd[1\\2\\3]\edge[1]%
\fwd[1\\2\\3]\edge[2]%
\fwd[1\\2\\3]\edge[\m1]%
\fwd[1\\2\\3]\edge[3]%
\fwd[1\\2\\3]\edge[1]%
\fwd[1\\2\\3]\edge[\m2]%
\fwd[1\\2\\3]\edge[\m1]%
\fwd[1\\2\\3]}.
\]
The U-traversal was introduced in two dimensions by Schrack and Liu~\cite{SL95}; the version presented here seems to be the most natural and straightforward way to generalize it to higher dimensions.

\paragraph{Gray-code traversal}\HowTo{ranking}
The Gray-code traversal~\cite{Sam06} follows the well-folded base pattern, and is reversed in every second subcube.
Example for $d = 3$ (see Figure~\ref{fig:discontinuous}):\[
\descr{%
\fwd[1\\2\\3]\edge[1]%
\rev[1\\2\\3]\edge[2]%
\fwd[1\\2\\3]\edge[\m1]%
\rev[1\\2\\3]\edge[3]%
\fwd[1\\2\\3]\edge[1]%
\rev[1\\2\\3]\edge[\m2]%
\fwd[1\\2\\3]\edge[\m1]%
\rev[1\\2\\3]}.
\]
We may call the traversal \emph{straight-jumping}: the coordinates of any pair of subcubes of equal size that are consecutive in the order (at any level of recursion) differ in exactly one dimension, but they are not necessarily adjacent.

\paragraph{Double-Gray-code traversal}\HowTo{row-coding, then ranking}
This traversal follows the well-folded base pattern. The traversal in the first subcube is a scaled-down version of the traversal of the unit cube. The traversals in the other subcubes are obtained from the traversal in the first subcube by reflections in the axis-parallel bisecting planes of the unit cube; moreover the traversal is reversed in every second subcube. Thus the transformation in the $i$-th subcube is $\big[-\mathring{c}_i[1]\cdot 1, -\mathring{c}_i[2]\cdot 2, ..., -\mathring{c}_i[d]\cdot d\big\}$ if $i$ is odd, and $\big\{-\mathring{c}_i[1]\cdot 1, -\mathring{c}_i[2]\cdot 2, ..., -\mathring{c}_i[d]\cdot d\big]$ if $i$ is even.
Example for $d = 3$ (see Figure~\ref{fig:discontinuous}):\[
\descr{%
\fwd[1\\2\\3]\edge[1]%
\rev[\m1\\2\\3]\edge[2]%
\fwd[\m1\\\m2\\3]\edge[\m1]%
\rev[1\\\m2\\3]\edge[3]%
\fwd[1\\\m2\\\m3]\edge[1]%
\rev[\m1\\\m2\\\m3]\edge[\m2]%
\fwd[\m1\\2\\\m3]\edge[\m1]%
\rev[1\\2\\\m3]}.
\]

This traversal is not only straight-jumping, but I found that it is also \emph{palindromic} as phrased by Bader~\cite{Bad13,Hav17}: if $A$ and $B$ are two subcubes sharing a $(d-1)$-dimensional facet $f$, then the subcubes of $B$ bordering $f$ (down to any depth of recursion) are visited in the opposite order of the adjacent subcubes of $A$ bordering $f$. This may make the traversal suitable for stack-and-stream algorithms for finite-element methods~\cite{Bad13,BMW06}.

The Double-Gray-code traversal was originally proposed by Faloutsos~\cite{Fal86} and described in more detail\footnote{In these descriptions, there may be some ambiguity regarding when to apply $g$ and when to apply $g^{-1}$. I~found that row-coding had to be done with $g$ and ranking with $g^{-1}$: replacing $g$ by $g^{-1}$ and/or vice versa would result in a traversal order that may still be consistent with Faloutsos's figure (which shows only the second level of refinement for $d=2$), but it would not be consistent with Samet's figure for the third level of recursion, and it would destroy the palindromic properties of the traversal.} by Samet~\cite{Sam06}.

\paragraph{Inside-out traversal}\HowTo{inversion, then row-coding, then ranking}
This traversal is similar to the Double-Gray-code traversal, but all subcubes are reflected in all dimensions to move the end points of the curves within the subcubes to the interior of the cube. Thus the transformation in the $i$-th subcube is $\big[\mathring{c}_i[1]\cdot 1, \mathring{c}_i[2]\cdot 2, ..., \mathring{c}_i[d]\cdot d\big\}$ if $i$ is odd, and $\big\{\mathring{c}_i[1]\cdot 1, \mathring{c}_i[2]\cdot 2, ..., \mathring{c}_i[d]\cdot d\big]$ if $i$ is even.
Example for $d = 3$ (see Figure~\ref{fig:discontinuous}):\[
\descr{%
\fwd[\m1\\\m2\\\m3]\edge[1]%
\rev[1\\\m2\\\m3]\edge[2]%
\fwd[1\\2\\\m3]\edge[\m1]%
\rev[\m1\\2\\\m3]\edge[3]%
\fwd[\m1\\2\\3]\edge[1]%
\rev[1\\2\\3]\edge[\m2]%
\fwd[1\\\m2\\3]\edge[\m1]%
\rev[\m1\\\m2\\3]}.
\]

The Inside-out traversal is a new design of my own. Its purpose is to improve the locality-preserving properties of the Double-Gray-code traversal, while maintaining all other potentially useful properties (such as palindromicity). Note that the traversal is still discontinuous: on the second level of recursion, consecutive subcubes are always adjacent (as shown in Figure~\ref{fig:discontinuous}), but on deeper levels of recursion, jumps between non-adjacent subcubes will appear.

\section{Simplex traversals}\label{sec:simplices}

The traversals in this section are based on subdividing simplices into $2^d$ simplices. More precisely, we consider the unit simplex defined by $\{x \in \Reals^d \mid \frac12 \geq x[1] \geq x[2] \geq ... \geq x[d] \geq -\frac12\}$; its vertices are the points $p_0,...,p_d$ whose coordinates are given by $p_h[j] = \frac12$ for $j \leq h$, and $p_h[j] = -\frac12$ for $j > h$. This simplex is a so-called Schl\"afli orthoscheme and a Hill simplex.

Bey~\cite{Bey92} and Burstedde and Holke~\cite{BuH16} defined enumeration or indexing schemes for recursive subdivisions of such simplices in two and three dimensions~\cite{BuH16}. Liu and Joe~\cite{LJ94} present a subdivision of a three-dimensional Hill ortoscheme into eight subsimplices\footnote{Liu and Joe start from another tetrahedron that is obtained from a Hill ortoscheme by bisection; a subdivision of a Hill ortoscheme into eight subsimplices is found on the second to fifth levels of their recursive subdivision scheme.}, but their description does not specify, for each subsimplex, which of the two possible similarity transformations from unit simplex to subsimplex should be applied. As a result, the traversal order is undefined. In higher dimensions, Freudenthal's subdivision scheme~\cite{Fre42} tiles a Hill simplex with $2^d$ mutually congruent smaller copies of itself. If the Hill simplex is also an orthoscheme, Maehara's bisection scheme~\cite{Mae14} results in an alternative subdivision into $2^d$ congruent smaller copies. However, I~do not know of any previous proposal to define the order in which to traverse the tiles for any number of dimensions. In this section, I propose two definitions of such traversal orders.

\paragraph{Hill-Z traversal}
This traversal defines an ordering on the subsimplices that are obtained with Freudenthal's subdivision scheme (explained in Inset~\ref{ins:whyHillZworks}).

Recall that in our notation, we take the centre of the simplex's bounding cube as the anchor point for its location. The signs of the coordinates of the $i$-th subsimplex's centre are obtained by sorting the binary digits of $i-1$ in non-increasing order, and interpreting $0$ as $-1$. The moves between them follow. For an example, see Table~\ref{tab:Hillpattern}.

\begin{table}
\begin{centering}
\caption{An example of how to obtain the base pattern of the Hill-Z traversal}\label{tab:Hillpattern}
\leavevmode\begin{tabular}{|r|cccccccc|}
\hline
$i$              & 1                & 2               & 3               & 4
                   & 5                & 6               & 7               & 8 \\
$i-1$            & 000              & 001             & 010             & 011
                   & 100              & 101             & 110             & 111 \\
$i-1$, reversed  & 000              & 100             & 010             & 110
                   & 001              & 101             & 011             & 111 \\
$i-1$, sorted    & 000              & 100             & 100             & 110
                   & 100              & 110             & 110             & 111 \\
$\mathring{c}_i$ & \vtx{-1\\-1\\-1} & \vtx{1\\-1\\-1} & \vtx{1\\-1\\-1} & \vtx{1\\1\\-1}
                   & \vtx{1\\-1\\-1}  & \vtx{1\\1\\-1}  & \vtx{1\\1\\-1}  & \vtx{1\\1\\1} \\
$V_i$            &                  & 1        & $\emptyset$     & 2
                   & $-2$        & 2        & $\emptyset$     & 3 \\
\hline
\end{tabular}\par
\end{centering}
\end{table}

\begin{figure}[b]
\centering
\includegraphics[width=\hsize,page=1]{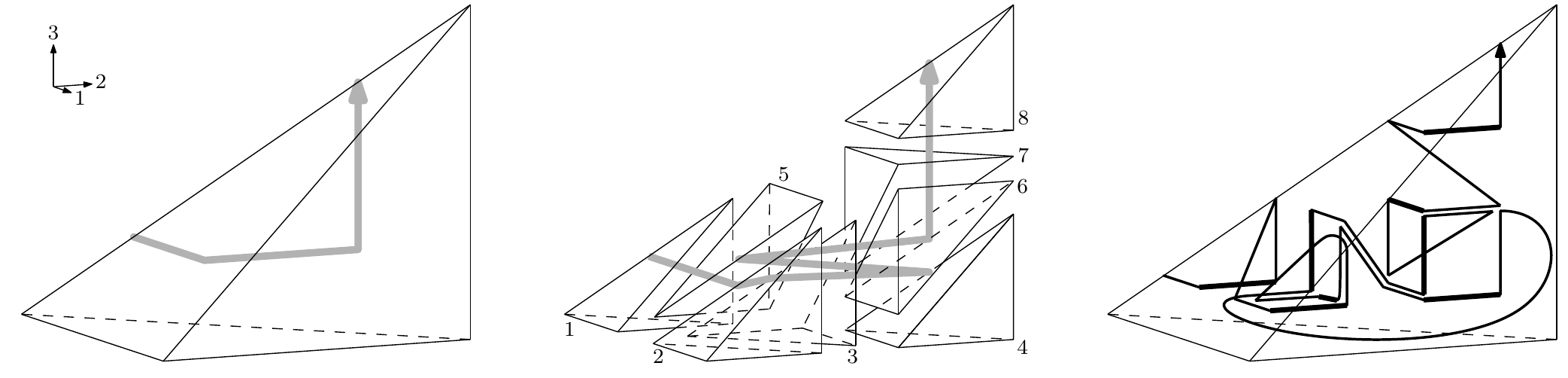}
\vspace{-12pt}
\caption{The Hill-Z traversal.}
\label{fig:hillztraversal}
\end{figure}

The transformation in subsimplex $i$ is described by the permutation of a stable sorting algorithm that would transform $i-1$, reversed, into $i-1$, sorted. More precisely, let $r[d],...,r[1]$ be the digits of the binary representation of $i-1$, in order from most significant to least significant. Then for each $j$, if $r[j] = 1$, then $\pi_i[j] = \sum_{h=1}^j r[h]$, and if $r[j] = 0$, then $\pi_i[j] = d - \sum_{h=j+1}^d (1-r[h])$. For example, for $d=3$ we get (see Figure~\ref{fig:hillztraversal}):\[
\descr{%
\fwd[1\\2\\3]\edge[1]%
\fwd[1\\2\\3]%
\fwd[2\\1\\3]\edge[2]%
\fwd[1\\2\\3]\edge[\m2]%
\fwd[2\\3\\1]\edge[2]%
\fwd[1\\3\\2]%
\fwd[3\\1\\2]\edge[3]%
\fwd[1\\2\\3]}.
\]

Note that there are no reflections or reversals. All simplices that share the same centre, are traversed starting at the bounding box vertex with minimum coordinates in all dimensions, and ending at the bounding box vertex with maximum coordinates in all dimensions---similar to Z-traversal. The Hill-Z traversal is discontinuous for any $d \geq 2$. Inset~\ref{ins:whyHillZworks} explains why the definition as presented above indeed leads to a space-filling traversal.

\begin{inset}
\caption{Why the Hill-Z traversal works}\label{ins:whyHillZworks}
\begin{wrapfigure}{r}{0.25\textwidth}
  \includegraphics[width=0.25\textwidth]{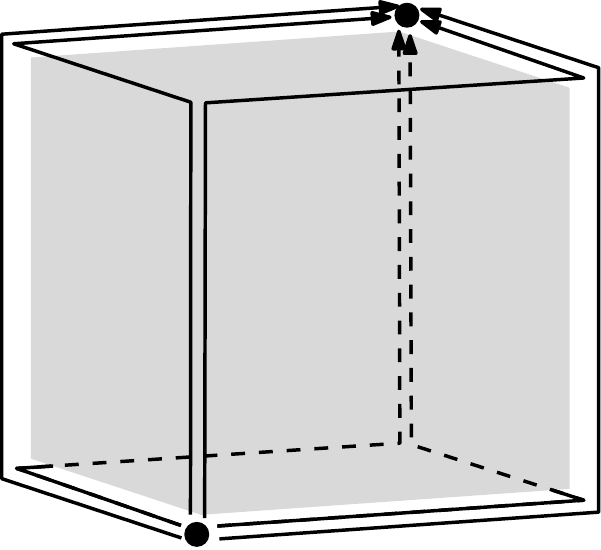}
  \vspace{-15pt}
  \caption{$d!$ paths from minimum to maximum.}
  \label{fig:sixpaths}
\end{wrapfigure}
Let $B$ be one of the $2^d$ subcubes of the unit cube. Let $y$ be the vertex of $B$ with minimum coordinates, and let $z$ be the vertex of $B$ with maximum coordinates. Any shortest path along the edges (one-dimensional faces) of $B$ from $y$ to $z$ contains an edge upwards along each of the $d$ dimensions of the cube. We can go up in the $d$ dimensions in any order, and thus there are $d!$ such paths; see Figure~\ref{fig:sixpaths} for an example. Observe that the convex hull of each such path is a Hill simplex, and together these $d!$ Hill simplices tile $B$.

Our unit simplex only intersects the subcubes that contain one of the vertices of the unit simplex. Suppose $B$ contains the vertex $p_h$. For any point in this subcube, we have $x_j \geq x_k$ for any $j \leq h < k$. The intersection of the unit simplex with $B$ contains only those tiles of $B$ that also satisfy the other constraints on $x$. Thus it contains only those tiles that are defined by a path in which the edges parallel to axes $1,...,h$ appear in order, and the edges parallel to axes $h+1,...,d$ appear in order. The number of such paths equals the number of ways in which a sequence of size $h$ and a sequence of size $d-h$ can be interleaved. This is exactly the number of ways in which a sequence of $h$ ones and $d-h$ zeros can be interleaved, that is, it is the number of $d$-digit binary numbers that contain exactly $h$ ones.

Our mapping therefore counts the number of ones in the $d$-digit binary number $i-1$ to determine $p_h$, and thus, the subsimplex centre point, and then maps the ones in the binary number $i-1$ to steps parallel to axes $1,...,h$, in order, and it maps the zeros in $i-1$ to steps parallel to axes $h+1,...,d$, in order.
\end{inset}

\paragraph{Maehara-reflected traversal}
I arrived at the design of the Maehara-reflected traversal in two independent ways. In this section I will describe both ways, because they give complementary insights in the structure of the traversal.

\textit{The Maehara-reflected traversal as a variation on the Hill-Z traversal.}
The subsimplex centre points of the Maehara-reflected traversal are the same as in the Hill-Z traversal, but reflections and reversals are used in an attempt to reduce discontinuities. In particular, going through the sequence of subsimplices, we will alternate between moving from a unit simplex vertex to the centre and moving from the unit simplex centre to a vertex.

\begin{figure}
\centering
\includegraphics[width=\hsize]{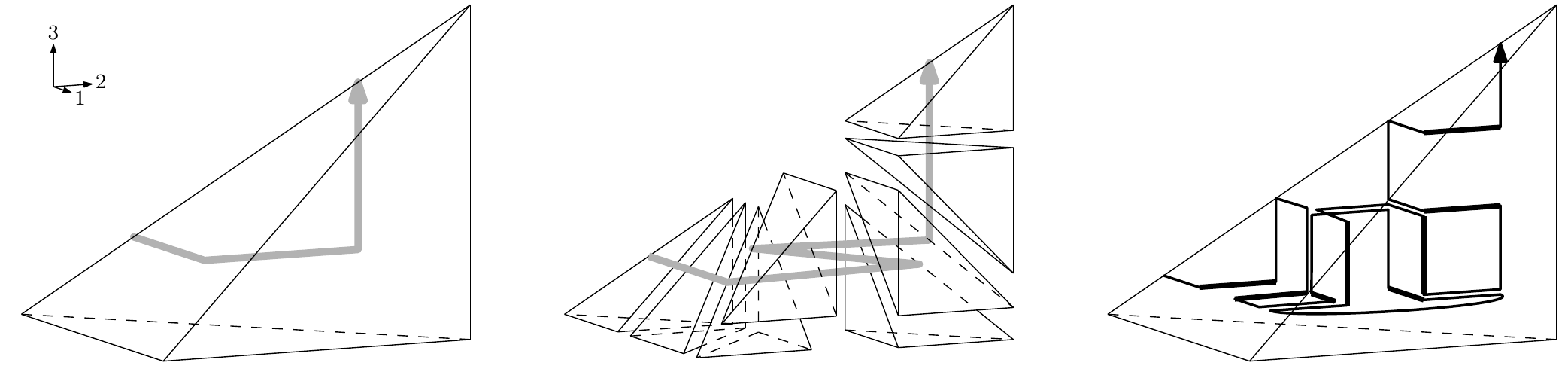}
\vspace{-12pt}
\caption{The Maehara-reflected traversal.}
\label{fig:maeharatraversal}
\end{figure}

Thus we get the following definition. Let $i$ be the index of a subsimplex, and let $r[d],...,r[1]$ be the digits of the binary representation of $i-1$, in order from most significant to least significant. The subsimplex's centre point $c_i$ is given by $c_i[j] = \frac14$ if $j \leq \sum_{h=1}^d r[h]$ and $c_i[j] = -\frac14$ if $j > \sum_{h=1}^d r[h]$. For each $j$, if $r[j] = 1$, then $\pi_i[j] = -\sum_{h=j}^d r[h]$, and if $r[j] = 0$, then, as with Hill-Z, we have $\pi_i[j] = d - \sum_{h=j+1}^d (1-r[h])$. Furthermore, the traversal is reversed in every second subsimplex, that is, if $r[1] = 1$. For example, for $d=3$ we get (see Figure~\ref{fig:maeharatraversal}):\[
\descr{%
\fwd[1\\2\\3]\edge[1]%
\rev[\m1\\2\\3]%
\fwd[2\\\m1\\3]\edge[2]%
\rev[\m2\\\m1\\3]\edge[\m2]%
\fwd[2\\3\\\m1]\edge[2]%
\rev[\m2\\3\\\m1]%
\fwd[3\\\m2\\\m1]\edge[3]%
\rev[\m3\\\m2\\\m1]}.
\]
Note that all subsimplices that have the same bounding cube are reflected in the same way, namely in the coordinates that are larger than zero. The reflections change the underlying recursive tessellation: it is no longer Freudenthal's tessellation, but it is the tessellation that is specific to Hill simplices that are also orthoschemes, as described by Maehara~\cite{Mae14}.

\break
\textit{The Maehara-reflected traversal as an extension of Maehara's bisection scheme.}
Maehara~\cite{Mae14} describes how, from a $d$-dimensional Hill orthoscheme $S_d$, we can derive $d$ simplices $S_0,...,S_{d-1}$ and $d$ simplices $T_0,...,T_{d-1}$ with the following properties. First, $S_0$ is $S$ scaled down by a factor two, with fixed point $p_0$. Second, $T_i$ is obtained from $S_i$ by a similarity transformation which, in our notation, reads $[-(i+1),...,-1,i+2,...,d\}$. Finally, $S_i$ and $T_i$ together tile $S_{i+1}$. If we choose to traverse each $T_i$ in the opposite order as compared to $S_i$, and add the implied moves, then this leads to the following definition of the $d$-dimensional Maehara-reflected traversal $H(d)$.

The one-dimensional traversal $H(1)$ is given by $\descr{\fwd[1]\edge[1]\rev[\m1]}$.
For $d > 1$, let $H'(d-1)$ be $H(d-1)$ with $d$ appended to each permutation; $H(d)$ is now defined as the concatenation of $H'(d-1)$, a move $\{-2,-3,...,-(d-1)\}$, and the reverse of $H'(d-1)$ subject to the transformation $[-d,...,-1\}$.

For example, $H'(1) = \descr{\fwd[1\\2]\edge[1]\rev[\m1\\2]}$,
its reverse is $\descr{\fwd[\m1\\2]\edge[\m1]\rev[1\\2]}$, which is subjected to the transformation $[-2,-1\}$
and thus:\begin{align*}
H(2)  &= \descr{\fwd[1\\2]\edge[1]\rev[\m1\\2]\fwd[2\\\m1]\edge[2]\rev[\m2\\\m1]}\\
H'(2) &= \descr{\fwd[1\\2\\3]\edge[1]\rev[\m1\\2\\3]\fwd[2\\\m1\\3]\edge[2]\rev[\m2\\\m1\\3]}\\
\mathrm{reverse}(H'(2)) &= \descr{\fwd[\m2\\\m1\\3]\edge[\m2]\rev[2\\\m1\\3]\fwd[\m1\\2\\3]\edge[\m1]\rev[1\\2\\3]}\\
\mathrm{reverse}(H'(2))\mbox{ subject to }\descr{\fwd[\m3\\\m2\\\m1]} &= \descr{\fwd[2\\3\\\m1]\edge[2]\rev[\m2\\3\\\m1]\fwd[3\\\m2\\\m1]\edge[3]\rev[\m3\\\m2\\\m1]}\\
\end{align*}
And thus:\[%
H(3) = \descr{\fwd[1\\2\\3]\edge[1]\rev[\m1\\2\\3]\fwd[2\\\m1\\3]\edge[2]\rev[\m2\\\m1\\3]\edge[\m2]
\fwd[2\\3\\\m1]\edge[2]\rev[\m2\\3\\\m1]\fwd[3\\\m2\\\m1]\edge[3]\rev[\m3\\\m2\\\m1]}.\]

\textit{Continuity properties.}
The one- and two-dimensional versions of the Maehara-reflected traversal are face-continuous. In fact, $H(2)$ is exactly the well-known P\'olya curve, also known as Sierpi\'nski-Knopp curve~\cite{Pol1913,Sie75} (see Figure~\ref{fig:trianglecurves}(a)), so the Maehara-reflected traversal may be considered a generalization of the P\'olya curve to higher dimensions. Note, however, that the three-dimensional version is not continuous: it is only semi-face-continuous (for a proof, see Inset~\ref{ins:maeharapfc}). Bader~\cite{Bad13} showed that in three dimensions, a truly face-continuous traversal based on the recursive tessellation used here is impossible. %Continuity properties for $d \geq 4$ have not been studied.

\begin{inset}
\caption{Proving that for $d=3$, the Maehara-reflected traversal is semi-face-continuous}\label{ins:maeharapfc}
In the Maehara-reflected traversal, for any $i \in \{2,...,8\}$, the first subsimplex of $C_i$, on any level of refinement, has a two-dimensional intersection with an earlier simplex, that is, a subsimplex $C_j$ with $j < i$. Likewise, for any $i \in \{1,...,7\}$, the last subsimplex of $C_i$, on any level of refinement, has a two-dimensional intersection with a later simplex, that is, a subsimplex $C_j$ with $j > i$. It follows by induction that any section of the traversal that starts at the beginning of the complete traversal or ends at the end of the complete traversal is face-continuous.

Now, given any segment $S = \bigcup_{a \leq t \leq b} \tau^{+}(t)$ of the traversal, we can zoom in on the smallest subsimplex $C'$ that contains the complete segment. Let $C'_1,...,C'_8$ be the subsimplices of $C'$. By definition, there are $i$ and $j$ such that $1 \leq i < j \leq 8$ and the segment's starting point $\tau^{+}(a)$ lies in $C'_i$ and the segment's end point $\tau^{+}(b)$ lies in $C'_j$. Using the above observations and the geometric arrangement of the subsimplices (see Figure~\ref{fig:maeharatraversal}), we can conclude that the interior of $S$ consists of at most two spatially connected components: one component that contains the intersection of $S$ with $C'_i$ and all complete subsimplices $C'_h$ such that $i < h < \min(5,j)$, and one component that contains the intersection of $S$ with $C'_j$ and all complete subsimplices $C'_h$ such that $\max(4,i) < h < j$. Thus the three-dimensional Maehara-reflected traversal is semi-face-continuous. (A similar line of argument was used by Burstedde to prove the semi-face-continuity of the Z-traversal~\cite{Bur15}.)
\end{inset}

\break
\section{Hilbert curves (continuous quadrant-by-quadrant traversals)}\label{sec:Hilbert}

In this section, I describe continuous, self-similar traversals that visit the $2^d$ subcubes of a cube one by one. For $d=1$, the trivial traversal of a line segment from one end to the other is the only traversal of this type. For $d=2$, the Hilbert curve~\cite{Hil1891} is the only traversal of this type~\cite{Hav17}. As a result, all of the traversals defined in this section are the same for $d=1$ and $d=2$ and can be considered a generalization of the Hilbert curve to higher dimensions (except for the Beta Hilbert curve, which is undefined for $d < 3$). For $d \geq 3$, all of the traversals presented in this section still follow the well-folded base pattern but they differ in the transformations within the subcubes.

\begin{figure}
\centering\leavevmode
\includegraphics[width=0.3\hsize,page=5]{traversals.pdf}\hfill
\includegraphics[width=0.3\hsize,page=1]{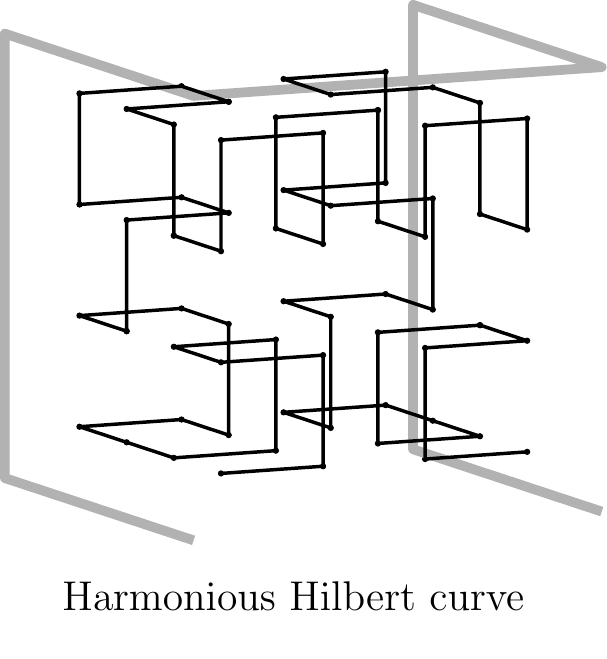}\hfill
\includegraphics[width=0.3\hsize,page=2]{traversals.pdf}\\[\baselineskip]
\includegraphics[width=0.3\hsize,page=4]{traversals.pdf}\quad
\raisebox{0.25\hsize}{\includegraphics[scale=1.3,page=19]{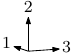}}\quad
\includegraphics[width=0.3\hsize,page=6]{traversals.pdf}
\caption{The three-dimensional versions of the space-filling curves in this paper that are based on subdividing cubes into $2^d$ subcubes. The Base-camp, Alfa and Beta Hilbert curves are asymmetric: the open dots in the figures indicate which subcube within each subcube is matched to the first subcube of the curve as a whole~\cite{Hav17}. The Base-camp curve follows a path through the first two subcubes to the centre of the unit cube (base camp), from where it makes $2^{d-1}-2$ excursions to pairs of subcubes, always returning to base camp, before it leaves the unit cube via the last two subcubes.}\label{fig:continuous}
\end{figure}

\paragraph{Base-camp Hilbert curve}
The Base-camp curve~\cite{Hav17} can be described as a next step in the progression from Z-traversal to Inside-out traversal in Section~\ref{sec:discontinuous}. In fact, the definition of the Base-camp curve differs from the Inside-out traversal only in the first and in the last subcube: in these subcubes, we swap the first and the last coordinate axis, and moreover, in the first subcube we reflect the traversal in all coordinates. Thus, the transformations $\gamma_i$ are given by $\gamma_1 = \big[d,2,...,d-1,1\big\}$; $\gamma_i = \big\{\mathring{c}_i[1]\cdot 1,...,\mathring{c}_i[d]\cdot d\big]$ for all even $i < 2^d$; $\gamma_i = \big[\mathring{c}_i[1]\cdot 1,...,\mathring{c}_i[d]\cdot d\big\}$ for all odd $i > 1$; and $\gamma_{D} = \big\{{-d},-2..,-(d-1),1\big]$. For example, for $d=3$ we get (see also Figure~\ref{fig:continuous}):\[
\descr{%
\fwd[3\\2\\1]\edge[1]%
\rev[1\\\m2\\\m3]\edge[2]%
\fwd[1\\2\\\m3]\edge[\m1]%
\rev[\m1\\2\\\m3]\edge[3]%
\fwd[\m1\\2\\3]\edge[1]%
\rev[1\\2\\3]\edge[\m2]%
\fwd[1\\\m2\\3]\edge[\m1]%
\rev[\m3\\\m2\\1]}.
\]

The result is a continuous traversal, that is, a curve, that, assuming a unit cube centred on the origin, starts at the unit cube vertex $(-\frac12,...,-\frac12)$ and ends in the centre of a $(d-2)$-dimensional face of the unit cube at $(-\frac12,0,...,0,\frac12)$.
In one dimension, we disambiguate $\gamma_{D}$ as $\{-1]$, which results in the trivial continuous traversal of a one-dimensional unit ``cube'' from one end to the other.

\break
\paragraph{Harmonious Hilbert curve}
The transformation within the $i$-th subcube of this curve is constructed as follows. Let $r[d],...,r[1]$ be the digits of the binary representation of $i-1$, from most significant to least significant. Start with the transformation $[d,...,1\}$ that reverses the order of all axes. In the sequence $\pi_i[1],...,\pi_i[d]$, move to the front all axes $j$ such that the $r[j] = r[1]$, while maintaining the order of these axes relative to each other. For each $j$, if $\pi_i[j] = 1$, give it the sign $\mathring{c}_i[1]$, and if $\pi_i[j] > 1$, give it the sign $-\mathring{c}_i[\pi_i[j]]$. Reverse the curve if $i$ is odd. %For example, in the six-dimensional curve, the seventh subcube ($i-1 = 000110$), with $\mathring{c}_7 = (1,-1,1,-1,-1,-1)$ gets transformation $\{6,5,4,1,-3,2]$.
For example, for $d = 3$ we get (see also Figure~\ref{fig:continuous}):\[
\descr{%
\rev[3\\2\\\m1]\edge[1]%
\fwd[1\\3\\2]\edge[2]%
\rev[3\\1\\\m2]\edge[\m1]%
\fwd[\m2\\\m1\\3]\edge[3]%
\rev[\m2\\\m1\\\m3]\edge[1]%
\fwd[\m3\\1\\\m2]\edge[\m2]%
\rev[1\\\m3\\2]\edge[\m1]%
\fwd[\m3\\2\\\m1]}.
\]
The result has the following characteristic property~\cite{Hav12}: the points on all $(d-1)$-dimensional facets of the unit cube, except one\footnote{The exception is the facet with midpoint $(\frac12,0,...,0)$.}, are visited in the order of an appropriately rotated \hbox{$(d-1)$}-dimensional Harmonious Hilbert curve.
%In particular, this implies that the $2d$-dimensional curve that starts in the origin of the coordinate system visits the points on the $d$-dimensional faces that contain the origin in the same order as the $d$-dimensional curve, which is the property we sought for its application in the construction of R-trees~\cite{HW11}.
Up to at least three dimensions, the Harmonious Hilbert curve is the only octant-by-octant self-similar continuous space-filling traversal with this property~\cite{Hav17}.
The Harmonious Hilbert curve is face-continuous.

\paragraph{Alfa and Beta Hilbert curves}
For both curves, permutations are constructed as follows. Start with $r[d],...,r[1]$, the digits of the binary representation of $i-1$, from most significant to least significant. Define $r'[-1,...,d]$ as follows:\[\begin{array}{lll}
r'[-1]  & = r[1]      & \\
r'[0]   & = 1-r[1]    & \\
r'[j]   & = r[j]      & \mbox{for $1 \leq j \leq d-2$} \\
r'[d-1] & = r[d-1]    & \mbox{if $i \leq (7/8)D$} \\
r'[d-1] & = 0         & \mbox{if $i > (7/8)D$, that is, $r[d] = r[d-1] = r[d-2] = 1$} \\
r'[d]   & = 1-r'[d-1] & \\
\end{array}\]
Now, for $j = 1,...,d$, let $\pi_i[j]$ be the smallest $h > d-j-1$ such that $r'[h] = r'[d-j-1]$. Reverse the direction of the traversal if $i$ is odd.

For the Alfa curve, continue as follows. Assign signs as with the Harmonious Hilbert curve: if $\pi_i[j] = 1$, give it the sign $\mathring{c}_i[1]$, and if $\pi_i[j] > 1$, give it the sign $-\mathring{c}_i[\pi_i[j]]$. Finally, apply the following corrections to all subcubes \emph{except} the first and the last: swap $\pi_i[d-1]$ and $\pi_i[d]$; change the sign of $\pi_i[d]$, and reverse the direction. For example, for $d = 3$ we get (see also Figure~\ref{fig:continuous}):\[
\descr{%                 -0123
\rev[2\\3\\\m1]\edge[1]%   01001
\rev[3\\1\\\m2]\edge[2]%   10101
\fwd[3\\1\\2]\edge[\m1]%  01010
\rev[\m2\\\m1\\\m3]\edge[3]%   10110
\fwd[\m2\\\m1\\3]\edge[1]%   01001
\rev[\m3\\1\\2]\edge[\m2]%  10101
\fwd[\m3\\1\\\m2]\edge[\m1]%  01010
\fwd[\m3\\2\\\m1]}.%         10101
\]

For the Beta curve, assign signs as with the Inside-out traversal: give $\pi_i[j]$ the sign $\mathring{c}_i[\pi_i[j]]$. Finally, apply the aforementioned corrections to \emph{only} the first and the last subcube. For example, for $d = 3$ we get (see also Figure~\ref{fig:continuous}):\[
\descr{%                 -0123
\fwd[\m2\\\m1\\3]\edge[1]%   01001
\fwd[\m3\\\m2\\1]\edge[2]%   10101
\rev[\m3\\2\\1]\edge[\m1]%  01010
\fwd[2\\\m3\\\m1]\edge[3]%   10110
\rev[2\\3\\\m1]\edge[1]%   01001
\fwd[3\\2\\1]\edge[\m2]%  10101
\rev[3\\\m2\\1]\edge[\m1]%  01010
\rev[3\\\m1\\2]}.%         10101
\]

%The Alfa curve can be considered a true generalization of the Hilbert curve, although in implementations it can be convenient to exploit conditions that are only satisfied in three or more dimensions. For example, the curve must be reversed in the first subcube and in every second subcube except the last, but the implementation of this condition in Line~26 of Algorithm~1\cite{BH16} is correct only for three- and higher-dimensional curves. The Beta curves are defined only for three or more dimensions: they share the defining characteristics of the two-dimensional Hilbert curve~\cite{Hav16}, but not vice versa.
The Alfa and Beta curves were found by Bos and myself: they are the self-similar, well-folded, \emph{hyperorthogonal} curves as described implicitly by Theorem 46 in our article~\cite{BH16}. The Alfa Hilbert curve is the version that starts and ends in vertices of the unit cube. It is well-defined for any $d \geq 1$ under a reasonable interpretation of its definition for the boundary cases $d = 1$ and $d = 2$. The Beta Hilbert curve is the version whose end points lie on $(d-1)$-dimensional facets of the unit cube, and is defined only for $d \geq 3$. The presentation above is the first attempt to define these curves in a direct way for any number of dimensions. One can verify that the above definition of $\pi_i[j]$ in terms of $r'$ implements the sorting of coordinate axes by ``local edge distance'' as in the original algorithm~\cite{BH16}.

Both curves are face-continuous. In three dimensions, the curves have excellent locality-preserving properties~\cite{Hav17}. In fact, regardless of the number of dimensions, one can show that every section of the curves has a bounding box whose volume is at most four times the volume of the curve section itself~\cite{BH16}.

\paragraph{The Butz-Hilbert curve}
Again, let $r[d],...,r[1]$ be the digits of the binary representation of $i-1$, from most significant to least significant. Let $k$ be the lowest index such that $r[k] \neq r[1]$; if such an index does not exist (because all digits are equal), take $k = 1$. Now start with $\Pi_i$ defined by $\pi_i[j] = j + k \pmod d$, and apply reflections (sign changes) and reversals as with the Harmonious Hilbert curve discussed above. For example, for $d = 3$ we get (see also Figure~\ref{fig:continuous}):\[
\descr{%
\rev[2\\3\\\m1]\edge[1]%
\fwd[3\\1\\2]\edge[2]%
\rev[3\\1\\\m2]\edge[\m1]%
\fwd[\m1\\\m2\\3]\edge[3]%
\rev[\m1\\\m2\\\m3]\edge[1]%
\fwd[\m3\\1\\\m2]\edge[\m2]%
\rev[\m3\\1\\2]\edge[\m1]%
\fwd[2\\\m3\\\m1]}.
\]
This generalization of the Hilbert curve, found by Butz~\cite{Btz71}, is probably the oldest generalization of the Hilbert curve to higher dimensions. Algorithms that implement Butz's curve can be found in various sources~\cite{Btz71,Law00,Moo00}. The Butz-Hilbert curve is face-continuous.

\section{Peano curves and variants}\label{sec:Peano}
\begin{figure}
\centering\leavevmode
\includegraphics[width=0.225\hsize,page=13]{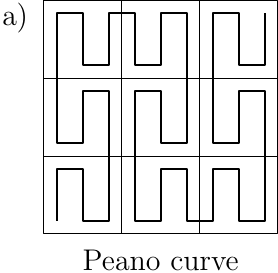}\hfill
\includegraphics[width=0.225\hsize,page=14]{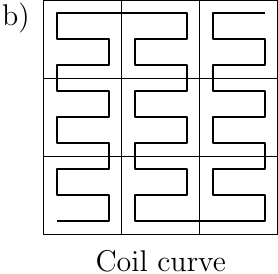}\hfill
\includegraphics[width=0.225\hsize,page=15]{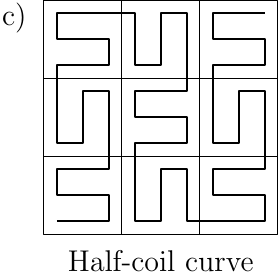}\hfill
\includegraphics[width=0.225\hsize,page=16]{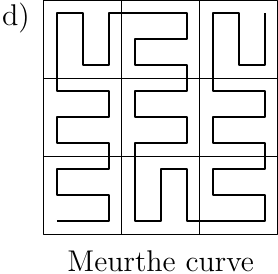}
\caption{The two-dimensional versions of the space-filling curves in this paper that are based on subdividing cubes into $3^d$ subcubes. All curves are order-preserving: they are defined without reversals.}\label{fig:ternarycurves}
\end{figure}

This section describes face-continuous, self-similar traversals that are based on subdividing cubes into $3^d$ subcubes (see Figure~\ref{fig:ternarycurves}) and whose two-dimensional versions attracted my attention due to their excellent locality-preserving properties~\cite{HW10}. The first curve discussed below is Peano's curve. Peano described his curve in two and three dimensions~\cite{Pea1890}. Different from Hilbert's curve, Peano's curve seems to have an obvious generalization to higher dimensions, as presented below. The two-dimensional Coil and Half-coil curves are well-known variations on Peano's curve (see, for example, Luxburg~\cite{Lux98}); the two-dimensional Meurthe curve is a lesser-known variation that was named and studied by Van Walderveen and myself~\cite{HW10}. Below I present generalizations of these curves to higher dimensions.

All of the curves presented in this section have the property that the points of each facet of the $d$-dimensional cube are traversed in the order of the corresponding $(d-1)$-dimensional curve (I prove this in a report~\cite{Hav12} that, unfortunately, antedates the development of the much more readable notation that we are using now).

All curves in this section follow the ternary reflected Gray code pattern $G_3(d)$. As a result, the curves described below all start and end in opposite corners of the unit cube, and therefore, reflections are constructed in the same way for all curves, as follows. Let $t_i[d],...,t_i[1]$ be the digits of the ternary representation of $i-1$, in order from most-significant to least-significant. We flip the signs of all $\pi_i[j]$ such that $t_i[\pi_i[j]]$ and $\sum_{h=1}^d t_i[h]$ have different parity. We do not use reversal in any of the curves described below---the only differences are in the actual (unsigned) permutations.

\paragraph{Peano curve}
In the Peano curves, all permutations (modulo the signs) are simply the identity permutation. For example, for $d = 3$ we get:\begin{gather*}
\descr{%
\fwd[1\\2\\3]\edge[ 1]\fwd[1\\\m2\\\m3]\edge[ 1]\fwd[1\\2\\3]\edge[ 2]%
\fwd[\m1\\2\\\m3]\edge[\m1]\fwd[\m1\\\m2\\3]\edge[\m1]\fwd[\m1\\2\\\m3]\edge[ 2]%
\fwd[1\\2\\3]\edge[ 1]\fwd[1\\\m2\\\m3]\edge[ 1]\fwd[1\\2\\3]\edge[ 3]%
}\\\descr{%
\fwd[\m1\\\m2\\3]\edge[\m1]\fwd[\m1\\2\\\m3]\edge[\m1]\fwd[\m1\\\m2\\3]\edge[\m2]%
\fwd[1\\\m2\\\m3]\edge[ 1]\fwd[1\\2\\3]\edge[ 1]\fwd[1\\\m2\\\m3]\edge[\m2]%
\fwd[\m1\\\m2\\3]\edge[\m1]\fwd[\m1\\2\\\m3]\edge[\m1]\fwd[\m1\\\m2\\3]\edge[ 3]%
}\\\descr{%
\fwd[1\\2\\3]\edge[ 1]\fwd[1\\\m2\\\m3]\edge[ 1]\fwd[1\\2\\3]\edge[ 2]%
\fwd[\m1\\2\\\m3]\edge[\m1]\fwd[\m1\\\m2\\3]\edge[\m1]\fwd[\m1\\2\\\m3]\edge[ 2]%
\fwd[1\\2\\3]\edge[ 1]\fwd[1\\\m2\\\m3]\edge[ 1]\fwd[1\\2\\3]%
}
\end{gather*}

\paragraph{Coil curve}
In the Coil curves, all permutations (modulo the signs) are the permutation $[d,d-1,...,1\}$ that reverses the order of the axes. For example, for $d = 3$ we get:\begin{gather*}
\descr{%
\fwd[3\\2\\1]\edge[ 1]\fwd[\m3\\\m2\\1]\edge[ 1]\fwd[3\\2\\1]\edge[ 2]%
\fwd[\m3\\2\\\m1]\edge[\m1]\fwd[3\\\m2\\\m1]\edge[\m1]\fwd[\m3\\2\\\m1]\edge[ 2]%
\fwd[3\\2\\1]\edge[ 1]\fwd[\m3\\\m2\\1]\edge[ 1]\fwd[3\\2\\1]\edge[ 3]%
}\\\descr{%
\fwd[3\\\m2\\\m1]\edge[\m1]\fwd[\m3\\2\\\m1]\edge[\m1]\fwd[3\\\m2\\\m1]\edge[\m2]%
\fwd[\m3\\\m2\\1]\edge[ 1]\fwd[3\\2\\1]\edge[ 1]\fwd[\m3\\\m2\\1]\edge[\m2]%
\fwd[3\\\m2\\\m1]\edge[\m1]\fwd[\m3\\2\\\m1]\edge[\m1]\fwd[3\\\m2\\\m1]\edge[ 3]%
}\\\descr{%
\fwd[3\\2\\1]\edge[ 1]\fwd[\m3\\\m2\\1]\edge[ 1]\fwd[3\\2\\1]\edge[ 2]%
\fwd[\m3\\2\\\m1]\edge[\m1]\fwd[3\\\m2\\\m1]\edge[\m1]\fwd[\m3\\2\\\m1]\edge[ 2]%
\fwd[3\\2\\1]\edge[ 1]\fwd[\m3\\\m2\\1]\edge[ 1]\fwd[3\\2\\1]%
}
\end{gather*}

\paragraph{Half-coil curve}
In the Half-coil curves, the permutations (modulo the signs) alternate between reversing the order of the axes if $i$ is odd (as in the Coil curves) and maintaining the order of the axes if $i$ is even (as in the Peano curve). For example, for $d = 3$ we get:\begin{gather*}
\descr{%
\fwd[3\\2\\1]\edge[ 1]\fwd[1\\\m2\\\m3]\edge[ 1]\fwd[3\\2\\1]\edge[ 2]%
\fwd[\m1\\2\\\m3]\edge[\m1]\fwd[3\\\m2\\\m1]\edge[\m1]\fwd[\m1\\2\\\m3]\edge[ 2]%
\fwd[3\\2\\1]\edge[ 1]\fwd[1\\\m2\\\m3]\edge[ 1]\fwd[3\\2\\1]\edge[ 3]%
}\\\descr{%
\fwd[\m1\\\m2\\3]\edge[\m1]\fwd[\m3\\2\\\m1]\edge[\m1]\fwd[\m1\\\m2\\3]\edge[\m2]%
\fwd[\m3\\\m2\\1]\edge[ 1]\fwd[1\\2\\3]\edge[ 1]\fwd[\m3\\\m2\\1]\edge[\m2]%
\fwd[\m1\\\m2\\3]\edge[\m1]\fwd[\m3\\2\\\m1]\edge[\m1]\fwd[\m1\\\m2\\3]\edge[ 3]%
}\\\descr{%
\fwd[3\\2\\1]\edge[ 1]\fwd[1\\\m2\\\m3]\edge[ 1]\fwd[3\\2\\1]\edge[ 2]%
\fwd[\m1\\2\\\m3]\edge[\m1]\fwd[3\\\m2\\\m1]\edge[\m1]\fwd[\m1\\2\\\m3]\edge[ 2]%
\fwd[3\\2\\1]\edge[ 1]\fwd[1\\\m2\\\m3]\edge[ 1]\fwd[3\\2\\1]%
}
\end{gather*}

\paragraph{Meurthe curve}
The permutation (modulo the signs) within the $i$-th subcube of this curve is constructed as follows. Let $t[d],...,t[1]$ be the digits of the ternary representation of $i-1$, from most significant to least significant. Start with the identity transformation $[1,...,d\}$. In the sequence $\pi_i[1],...,\pi_i[d]$, move to the back all axes $j$ such that $t_i[j] \in \{0,1\}$, while reversing the order of these axes relative to each other. For example, for $d = 3$ we get:\begin{gather*}
\descr{%
\fwd[3\\2\\1]\edge[ 1]\fwd[\m3\\\m2\\1]\edge[ 1]\fwd[1\\3\\2]\edge[ 2]%
\fwd[\m3\\2\\\m1]\edge[\m1]\fwd[3\\\m2\\\m1]\edge[\m1]\fwd[\m1\\\m3\\2]\edge[ 2]%
\fwd[2\\3\\1]\edge[ 1]\fwd[\m2\\\m3\\1]\edge[ 1]\fwd[1\\2\\3]\edge[ 3]%
}\\\descr{%
\fwd[3\\\m2\\\m1]\edge[\m1]\fwd[\m3\\2\\\m1]\edge[\m1]\fwd[\m1\\3\\\m2]\edge[\m2]%
\fwd[\m3\\\m2\\1]\edge[ 1]\fwd[3\\2\\1]\edge[ 1]\fwd[1\\\m3\\\m2]\edge[\m2]%
\fwd[\m2\\3\\\m1]\edge[\m1]\fwd[2\\\m3\\\m1]\edge[\m1]\fwd[\m1\\\m2\\3]\edge[ 3]%
}\\\descr{%
\fwd[3\\2\\1]\edge[ 1]\fwd[\m3\\\m2\\1]\edge[ 1]\fwd[1\\3\\2]\edge[ 2]%
\fwd[\m3\\2\\\m1]\edge[\m1]\fwd[3\\\m2\\\m1]\edge[\m1]\fwd[\m1\\\m3\\2]\edge[ 2]%
\fwd[2\\3\\1]\edge[ 1]\fwd[\m2\\\m3\\1]\edge[ 1]\fwd[1\\2\\3]%
}
\end{gather*}

\section{Traversals that have not been generalized}\label{sec:lowd}

This section discusses several self-similar traversals that have been defined in at most three dimensions, but have not been generalized to higher dimensions yet. Clearly there is an overwhelming number of such traversals, many of which are rather arbitrary, and we cannot discuss them all. Therefore we restrict the discussion to a selection of traversals with fairly unique properties.

\paragraph{The P\'olya/Sierpi\'nski curve}
A two-dimensional Hill orthoscheme, that is, an isosceles right triangle, can be filled with the well-known P\'olya curve, also known as Sierpi\'nski-Knopp curve~\cite{Pol1913,Sie75} (see Figure~\ref{fig:trianglecurves}(a)). The two-dimensional Maehara-reflected traversal, presented in Section~\ref{sec:simplices}, is, in fact, the P\'olya curve. However, I would be reluctant to claim that the Maehara-reflected traversal is the most natural way to generalize the P\'olya curve, because the higher-dimensional versions are not continuous.
%A satisfactory generalization of the P\'olya curve to higher dimensions may actually be impossible: none of the known tetrahedra that can be tiled by eight smaller copies of themselves, admit a face-continuous traversal.

\paragraph{Palindromic tetrahedral traversal}
Van der Plas~\cite{Pla16} shows that the following discontinuous traversal of the three-dimensional Hill orthoscheme has the palindromic properties that may make it suitable for stack-and-stream processing~\cite{Bad13,BMW06}---and as such, it could be considered an alternative generalization of the P\'olya curve:\[\descr{%
\fwd[1\\2\\3]\edge[1]
\rev[\m1\\2\\3]
\fwd[2\\\m1\\3]\edge[2]
\rev[\m2\\\m1\\3]
\fwd[\m2\\3\\\m1]\edge[\m2]
\rev[2\\3\\\m1]\edge[2\\3]
\fwd[\m3\\\m2\\\m1]\edge[\m3]
\rev[3\\\m2\\\m1]}
\]
The traversal follows Maehara's tessellation. I do not know of a generalization of this traversal to higher dimensions.

%\paragraph{Bey's traversal}
%Bey~\cite{Bey92} described an enumeration scheme that amounts to the following self-similar traversal when applied to three-dimensional Hill ortoschemes:\[\descr{%
%\fwd[1\\2\\3]\edge[1]
%\fwd[1\\2\\3]\edge[2]
%\fwd[1\\2\\3]\edge[3]
%\fwd[1\\2\\3]\edge[\m2\\\m3]
%\fwd[2\\3\\1]
%\fwd[2\\1\\3]\edge[2]
%\fwd[3\\1\\2]
%\fwd[1\\3\\2]
%}\]
%The traveral follows the tessellation for general Hill simplices. The traversal can be roughly described as first covering the four subsimplices in the corners of the unit simplex, before going into the remaining subsimplices in the interior of the unit simplex.

\paragraph{SUB$_8$ traversal}
Liu and Joe~\cite{LJ96} describe a method to subdivide a tetrahedron into eight subtetrahedra. When the original tetrahedron is a type-1 Sommerville tetrahedron~\cite{Som23} (the Hill tetrahedron with four edges of length $\sqrt 3$ and two edges of length $2$), then the enumeration of vertices of each subsimplex in Liu and Joe's description can be interpreted as a similarity transformation from the unit simplex to the subsimplex. Together with the order in which the subtetrahedra are listed, this specifies a discontinuous traversal that could be specified as follows. Assuming a unit simplex with vertices $(-\frac12,-1,0)$, $(\frac12,0,-1)$, $(-\frac12,1,0)$, and $(\frac12,0,1)$, we define the centre points and the permutations of the subsimplices by:

\bgroup
\def\arraystretch{1.5}
\medskip
\begin{centering}\leavevmode\begin{tabular}{@{}c|cccccccc@{}}
$i$     & 1 & 2 & 3 & 4 & 5 & 6 & 7 & 8 \\
$c_i$   &
$(-\frac14,-\frac12,0)$ &
$(\frac14,0,-\frac12)$ &
$(\frac14,0,\frac12)$ &
$(-\frac14,\frac12,0)$ &
$(0,0,-\frac14)$ &
$(0,\frac14,0)$ &
$(0,-\frac14,0)$ &
$(0,0,\frac14)$ \\
$\pi_i$ & [1,2,3\} & [1,2,3\} & [1,2,3\} & [1,2,3\} & [3,2,-1\} & [-2,3,1\}& [2,3,1\} &
[-3,2,-1\}
\end{tabular}\end{centering}
\egroup
\medskip

\noindent There are no reversals.
Note that the centre points of the subsimplices in this traversal do not correspond to the intersection of a convex shape with a standard Cartesian grid. The traversal can be roughly described as first covering the four subsimplices in the corners, without any rotations or reflections, before going into the remaining subsimplices in the interior.

\paragraph{Various three-dimensional Hilbert curves}
My article on three-dimensional Hilbert curves contains an in-depth discussion of octant-by-octant self-similar cube-filling curves~\cite{Hav17}.  Several of these could be a starting point for alternative generalizations of the Hilbert curve to higher dimensions, with, for example, a more regular structure (so-called \emph{metasymmetric} curves), or different, non-well-folded, base patterns.

\paragraph{The Meander curve}
\begin{wrapfigure}{r}{0.225\textwidth}
  \vspace{-11pt}
  \includegraphics[width=0.225\textwidth,page=18]{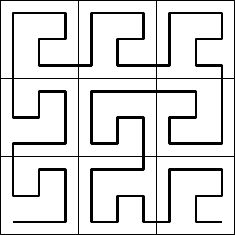}
  \vspace{-15pt}
  \caption{The Meander curve.}
  \label{fig:meander}
\end{wrapfigure}
A traversal of a square that recursively visits its subsquares in a grid of $3 \times 3$, can be face-continuous if the subsquares are visited in one of two patterns. One pattern, called ``serpentine'' by Wunderlich~\cite{Wun73}, underlies all curves presented in Section~\ref{sec:Peano}. The other pattern, which Wunderlich called ``meander'', underlies the curve sketched in Figure~\ref{fig:meander}. One may choose for each subsquare whether to use a forward traversal, or a reversed copy, so that there are $2^9 = 512$ ways to construct a self-similar face-continuous curve from this pattern. Most commonly we find the forward-only version shown here, defined by:\[
\descr{%
\fwd[2\\1]\edge[1]\fwd[2\\1]\edge[1]\fwd[1\\2]\edge[2]\fwd[1\\2]\edge[2]\fwd[1\\2]\edge[\m1]\fwd[\m1\\\m2]\edge[\m2]\fwd[\m2\\\m1]\edge[\m1]\fwd[\m2\\\m1]\edge[2]\fwd[1\\2]%
}
\]
I do not know of a generalization of this curve to higher dimensions. The first challenge in designing such a generalization would be to define the characteristics of this curve in a way that is independent of the number of dimensions.

\break
\section{Squared traversals}\label{sec:squared}

As explained in Section~\ref{sec:mapping}, any space-filling traversal can be regarded as a mapping $\tau$ from $[0,1]$ to $C$ (if the traversal is discontinuous, we would need to define whether to use $\tau^-$ or $\tau^+$, or another tie-breaking mechanism). If $C$ is a $d$-dimensional unit cube, and we translate it to $[0,1]^d$, then, for any $t \in [0,1]$, we can apply $\tau$ to each of the coordinates of $\tau(t)$ again. Thus we turn each of the $d$ coordinates of $\tau(t)$ into $d$ coordinates, so that we end up with $d^2$-dimensional points. More precisely, if $\tau(t)$ traverses the $d$-dimensional unit cube $[0,1]^d$ as $t$ goes from 0 to~1, then we can define a traversal $\tau'$ of the $d^2$-dimensional unit cube by $\tau'(t)[(i-1)d+j] = \tau(\tau(t)[i])[j]$ for all $i, j \in \{1,...,d\}$. Some properties of $\tau$ carry over to $\tau'$, in particular continuity and subcube structure, but self-similarity is not always preserved, as I will explain in this section.

\paragraph{Continuity} If the original traversal $\tau$ is continuous, then so is the squared traversal $\tau'$ (for a proof, see Sagan~\cite{Sag94}, Section 6.9, where the method is attributed to Steinhaus~\cite{Sth36}).

\paragraph{Subcube structure}
If the original traversal $\tau$ traverses the $s^d$ subcubes of the $d$-dimensional unit cube one by one, then the squared traversal $\tau'$ traverses the $s^{d\cdot d}$ subcubes of the $d^2$-dimensional unit cube one by one. This can be seen as follows. Consider the $d$-th level of refinement of $\tau$, where intervals of size $s^{-d\cdot d}$ map to $d$-dimensional subcubes of width $s^{-d}$. Each coordinate of such a subcube maps, in turn, to a $d$-dimensional subcube of width $s^{-1}$. As a result, $\tau'$ maps intervals of size $s^{-d \cdot d}$ to the Cartesian product of $d$ $d$-dimensional cubes of width~$s^{-1}$.

\paragraph{Self-similarity}
Consider the unit interval subdivided into $k^{d \cdot d}$ subintervals, and consider any one of these intervals, say $I = [(i-1)/k^{d \cdot d}, i/k^{d \cdot d}]$. The original traversal $\tau$ maps $I$ to the Cartesian product $X$ of $d$ subintervals $[(x[1]-1)/k^d, x[1]/k^d], [(x[2]-1)/k^d, x[2]/k^d], ..., [(x[d]-1)/k^d, x[d]/k^d]$, which is subsequently mapped to one of the $k^{d\cdot d}$ subcubes of the $d^2$-dimensional unit cube. The order in which this subcube $C'$ is traversed now depends on the following aspects of $\tau$: (i) the base pattern; (ii) the \emph{top-level transformations}, that is, the permutations, reflections and possible reversal that apply to $X$, the $i$-th subcube in the $d$-th level of subdivision; (iii) the \emph{low-level transformations}, that is, the permutations, reflections, and possible reversals that apply to the $x[j]$-th subcube of $\tau$, for $j = 1,...,d$.

What does it take for the traversal of $C'$ to be similar to the traversal of the full $d^2$-dimensional unit cube? Note that the low-level permutations and reflections directly induce permutations and reflections of some of the $d^2$ dimensions in the image of $\tau'$: thus they constitute symmetries of the $d^2$-dimensional cube. Furthermore, the possible top-level reversal would simply induce a complete reversal of the traversal of $C'$. Thus, these transformation do not affect the similarity between the traversals in different subcubes.

The effect of the other transformations, however, can be more complicated. To start with, a low-level reversal is not a reversal. It affects only $d$ of the $d^2$ dimensions, and does not, in general, constitute a symmetry of the $d$-dimensional unit cube. A top-level reflection has the same effect, as it results in using the reverse of a subtraversal of $\tau$ instead of the original, forward, subtraversal of $\tau$ to expand one coordinate to $d$ coordinates. For an example, see Appendix~\ref{apx:notselfsimilar}.

Suppose, however, that $\tau$ is symmetric, that is, equal to its own reverse under a similarity transformation that is a symmetry of the $d$-dimensional unit cube. Then a low-level reversal, and therefore also a top-level reflection, can be modelled as a simple low-level rotation and/or reflection in $d$ of the $d^2$ coordinates. Thus the traversals of all $d^2$-dimensional subcubes $C'$ that have the same top-level (unsigned) permutation are similar to each other, regardless of the top-level reflections, top-level reversals and low-level transformations that apply. The top-level permutations affect which coordinate of the original $d$-dimensional traversal is turned into which group of $d$ coordinates by which (low-level) transformation of the $d$-dimensional curve. Since similarity is preserved regardless of the low-level transformations, all traversals of all $d^2$-dimensional subcubes $C'$ are similar to each other and to the traversal of the full $d^2$-dimensional unit cube.

Specifically, given a representation of a symmetric traversal $\tau$ without reversals, the transformation $\pi'$ of $\tau'$ in $C'$ can be calculated as follows. Let $\sigma$ be the signed permutation that maps the reverse of $\tau$ to $\tau$, and let $\pi$ be the signed permutation that maps $\tau$ to the traversal of $X = [(x[1]-1)/k^d, x[1]/k^d] \times [(x[2]-1)/k^d, x[2]/k^d] \times ... \times [(x[d]-1)/k^d, x[d]/k^d]$ (modulo scaling and translation). Define $a \oplus b$ as the operator that adds $|b|$ to $a$ and gives the result the sign of $b$, that is, $a \oplus b = \mathring{b}(a + |b|)$. Then, for $j, j' \in \{1,...,d\}$, we have $\pi'[(j-1)d + j'] = (|\pi[j]|-1)d \oplus \pi_{x[|\pi[j]|]}[j']$ if $\pi[j]$ is positive, and $\pi'[(j-1)d + j'] = (|\pi[j]|-1)d \oplus (\pi_{x[|\pi[j]|]}\cdot\sigma)[j'] $ if $\pi[j]$ is negative.

Thus, self-similarity and symmetry of $\tau$ constitute sufficient conditions for the self-similarity of $\tau'$. The question what conditions are necessary, remains to be answered.

\paragraph{Examples of squared curves}
In the case of the Z-traversal, the Double-Gray-code traversal, and the Peano curve, squaring the traversals does not seem to bring anything new: the squared 2-dimensional traversal appears to be the same as the directly defined 4-dimensional traversal as described in Sections \ref{sec:discontinuous} and~\ref{sec:Peano}. In other cases, the squaring technique offers an alternative approach to generalizing low-dimensional curves to higher dimensions. Squaring the U-traversal, the Inside-out traversal, the Hilbert curve, the Coil curve, and the Half-coil curve results in self-similar traversals, since these traversals are symmetric. The squared two-dimensional Inside-out traversal is as follows:\begin{gather*}\descr{%
\fwd[1\\2\\3\\4]\edge[\m1]
\fwd[\m1\\2\\3\\\m4]\edge[\m3]
\fwd[\m1\\2\\\m3\\4]\edge[1]
\fwd[1\\2\\\m3\\\m4]\edge[2]
\fwd[1\\\m2\\\m3\\4]\edge[\m1]
\fwd[\m1\\\m2\\\m3\\\m4]\edge[3]
\fwd[\m1\\\m2\\3\\4]\edge[1]
\fwd[1\\\m2\\3\\\m4]\edge[4]}\\\descr{%
\fwd[1\\\m2\\3\\\m4]\edge[\m1]
\fwd[\m1\\\m2\\3\\4]\edge[\m3]
\fwd[\m1\\\m2\\\m3\\\m4]\edge[1]
\fwd[1\\\m2\\\m3\\4]\edge[\m2]
\fwd[1\\2\\\m3\\\m4]\edge[\m1]
\fwd[\m1\\2\\\m3\\4]\edge[3]
\fwd[\m1\\2\\3\\\m4]\edge[1]
\fwd[1\\2\\3\\4]
}.\end{gather*}
The squared Hilbert curve is the following:\begin{gather*}\descr{%
\fwd[2\\1\\4\\3]\edge[3]
\fwd[3\\4\\2\\1]\edge[1]
\fwd[3\\4\\1\\2]\edge[\m3]
\fwd[1\\\m2\\4\\\m3]\edge[2]
\fwd[4\\3\\1\\2]\edge[\m1]
\fwd[\m2\\\m1\\4\\3]\edge[3]
\fwd[\m2\\\m1\\3\\4]\edge[1]
\fwd[3\\\m4\\1\\\m2]\edge[4]}\\\descr{%
\fwd[3\\4\\1\\2]\edge[\m1]
\fwd[\m2\\\m1\\3\\4]\edge[\m3]
\fwd[\m2\\\m1\\\m4\\\m3]\edge[1]
\fwd[\m4\\3\\1\\\m2]\edge[\m2]
\fwd[1\\\m2\\\m4\\3]\edge[3]
\fwd[3\\\m4\\1\\\m2]\edge[\m1]
\fwd[3\\\m4\\2\\\m1]\edge[\m3]
\fwd[2\\1\\\m4\\\m3]
}.\end{gather*}

Squaring the Gray-code traversal, the Meurthe curve, and the Meander curve, which are all asymmetric, results in non-self-similar traversals.

\section{Implementation}\label{sec:practical}

Source code in C++ for two prototype software tools is available (please check my website or ask me for a copy by e-mail).

\paragraph{Generating definitions}
The first tool, \texttt{describe-traversal}, takes two arguments on its command line: the traversal type (one of the 16 discussed in Sections \ref{sec:discontinuous} through~\ref{sec:Peano}) and the number of dimensions. The output is a definition of the traversal, using the notation from Section~\ref{sec:definitionbypermutations}, produced by implementing the definitions as described in the previous sections.

\paragraph{Generating traversals}
The second tool, \texttt{generate-path}, takes as input the definition of a traversal in the notation from Section~\ref{sec:definitionbypermutations}, and up to three arguments on the command line:\begin{enumerate}
\item the desired level of refinement;
\item the exponent: 1~for a simple traversal (default), 2~for a squared traversal (higher exponents have not been implemented at this point);
\item where to put the origin of the coordinate system: at the centre of the unit cube, at the lexicographically smallest corner, at the centre of the first lowest-level subcube visited, or at the centre of the last lowest-level subcube visited.
\end{enumerate}
The output consists of the centre points of the lowest-level subcubes or subsimplices, in the order in which they are visited by the traversal. The coordinates are scaled so that they can all be expressed as integers.

The underlying algorithm is effectively the following. Recall from Section~\ref{sec:definitionbypermutations} that our notation is devised as a concise way of specifying, for each subcube $C_i$, the transformation matrix $M_i$ that needs to be applied, its centre point $c_i$, and the direction of the traversal---from now on we will denote the direction by $h_i$, where $h_i = 1$ for a forward traversal and $h_i = -1$ for a reversed traversal. Let $s$ be the scale factor, that is, the width of the unit cube divided by the width of a subcube.

The following algorithm extends a list of points, $P$, with the traversal of a cube or simplex centred on the point $c$, scaling, rotating and/or reflecting the traversal according to the transformation matrix $M$, going in the direction specified by $h$ (forward is~1, backward is~$-1$), refined to a depth of $r$ levels:

\SetFuncSty{textsc}
\SetProcNameSty{textsc}
\SetDataSty{textit}
\SetCommentSty{textrm}
\DontPrintSemicolon
\SetKw{KwDown}{down}
\setlength\algomargin{0em}
\SetAlCapHSkip{0em}
\begin{procedure}[H]
\caption{generatePath(list $P$, point $c$, matrix $M$, direction $h$, depth $r$)}
\If{$r = 0$}{Append $c$ to $P$}%
\ElseIf{$h = 1$}{\lFor{$i \gets 1$ \KwTo $s^d$}{\generatePath($P$, $c + M c_i$, $M M_i / s$, $h \cdot h_i$, $r-1$)}}%
\Else{\lFor{$i \gets s^d$ \KwDown \KwTo $1$}{\generatePath($P$, $c + M c_i$, $M M_i / s$, $h \cdot h_i$, $r-1$)}}
\end{procedure}

To get the full path for recursion depth $r$, taking the lexicographically smallest corner of the bounding box as the origin of the coordinate system, we run:

\begin{procedure}[H]
\caption{generateFullPath(depth $r$)}
$P \gets$ empty list\;
$T \gets$ $d$-dimensional matrix with $T[i,j] = s^r$ if $i = j$ and $T[i,j] = 0$ if $i \neq j$\;
\ProcNameSty{generatePath}($P$, 0, $T$, 1, $r$)\;
$p \gets$ a $d$-dimensional vector\;
\lFor{$j \gets 1$ \KwTo $d$}{$p[j] \gets$ the minimum $j$-th coordinate of all points in $P$}
Subtract $p$ from all points of $P$\;
\Return $P$\;
\end{procedure}

For a squared traversal, we run:

\begin{procedure}[H]
\caption{squaredTraversal(depth $r$)}
$X[0..(D^r-1)] \gets$ \ProcNameSty{generateFullPath}($r$)\;
$Q \gets$ \ProcNameSty{generateFullPath}($dr$)\;
$P \gets$ empty list\;
\For{$q \gets$ all points of $Q$, in order}{%
  $p \gets$ a $d^2$-dimensional vector\;
  \For{$i \gets 1$ \KwTo $d$}{\lFor{$j \gets 1$ \KwTo $d$}{$p[(i-1)d+j] \gets X[q[i]][j]$}}
  Append $p$ to $P$\;
}
\Return $P$\;
\end{procedure}

\paragraph{Adapting the space-filling curves to other shapes}
The cube-filling curves can also be used to fill parallelotopes, using an affine transformation to convert between cubes and parallelotopes. All subcubes' corresponding parallelotopes will have the same shape. Similarly, the Hill-orthoscheme-filling curves can be used to fill arbitrary simplices, using an affine transformation to convert between the different shapes. However, in that case, subsimplices may get different geometric shapes, that is, angles may change, depending on how the subsimplices are rotated and reflected in the original grid of mutually congruent orthoschemes. With traversals based on Freudenthal's subdivision scheme, such as Hill-Z, there can be up to $d!$ different shapes (only one if the target simplex is still a Hill simplex). With the Maehara-reflected traversals the number of different shapes may become much higher, but it may very well be possible to prove lower bounds on the quality of those shapes if used as elements of a computational mesh; Liu and Joe~\cite{LJ94} studied the three-dimensional case.

\section{Unanswered questions and work to do}\label{sec:conclusions}

\paragraph{Implementations}
To be able to put traversals to use in practical applications, we typically need to implement a few more tools, for example to:\begin{itemize}
\item generate actual subsimplices rather than mere anchor points;
\item compute the position of a point in the traversal (this may require a tie-breaking mechanism, for example, given a point $p \in \Reals^d$, compute the smallest $t$ such that $\tau^+(t) = p$);
\item given a set of points $S$ in $d$-dimensional space, compute a set of curve sections that (approximately) covers $S$;
\item compute, given two points in $d$-dimensional space, which of the two is visited first by the traversal;
\item find the neighbours of a subsimplex in $d$-dimensional space.
\end{itemize}
For the cube-based traversal, most of this is relatively simple; for simplex-based traversals, it is more involved. Most of these operations would require an algorithm that uses an explicit representation of the shape of the unit simplex. This is not trivial to calculate automatically from the definition of the traversal, so we might have to extend our notation to encode this. For three-dimensional Hill orthoschemes, it would be good to investigate whether there are substantial differences in ease and efficiency of implementation between the Hill-Z traversal, the three-dimensional traversal from Burstedde and Holke~\cite{BuH16}, the Maehara-reflected traversal, and the palindromic tetrahedral traversal~\cite{Pla16}. (Note that the first two follow Freudenthal's subdivision scheme, whereas the last two follow Maehara's subdivision scheme.)

\paragraph{Composite traversals}
It may also be useful to extend our notation scheme and the accompanying software to deal with ``composite'' traversals that are defined by multiple refinement rules that refer to each other. The three-dimensional simplex traversal from Burstedde and Holke~\cite{BuH16} is of that type, as is the continuous (but not face-continuous) three-dimensional simplex traversal by Bader~\cite{Bad13}\footnote{Section ``A three-dimensional, node-connected quasi-Sierpinski curve''~\cite{Bad13}.}. Noteworthy composite traversals of squares include Wierum's $\beta\Omega$-curve~\cite{Wie02}, the $AR^2W^2$-curve~\cite{ARRWW97} and the Kochel curve~\cite{Hav11}. The $\beta\Omega$-curve improves~\cite{HW10,Wie02} on the locality-preserving properties of Hilbert's curve by crossing boundaries between squares in the interior of their edges rather than at vertices. The $AR^2W^2$-curve and the Kochel curve are designed to minimize the maximum number of curve sections that are needed to adequately cover any smaller square (not limited to squares that appear in the recursive subdivision). Furthermore, being able to deal with composite traversals would allow us to write down a description of any squared traversal directly in $d^2$ dimensions, so that any tool can process it in the same way as non-squared traversals.

\paragraph{Alternative and further generalizations}
In three dimensions, there are several octant-by-octant self-similar space-filling curves that may still be developed into alternative generalizations of the Hilbert curve to any number of dimensions~\cite{Hav17}. The composite square-filling curves mentioned above have not been generalized to higher dimensions either. However, one may regard the Beta curves presented in Section~\ref{sec:Hilbert} as a generalization of the $\beta\Omega$-curve\footnote{The Beta curves are, in fact, named after the $\beta$-pattern of Wierum's $\beta\Omega$-curve, and the Alfa curves were subsequently named to match the Beta curves.}, and I~found that, in higher dimensions, the cube-by-cube traversals of Sections \ref{sec:discontinuous} and~\ref{sec:Hilbert} must already have near-optimal performance~\cite{Hav11} according to the design objectives of the $AR^2W^2$-curve and the Kochel curve.

Clearly, the bigger gaps in our knowledge are about traversals of other shapes than cubes.

In this paper, we described traversals based on Maehara's subdivision scheme on Hill orthoschemes; Freudenthal's subdivision scheme on Hill simplices; and Liu and Joe's subdivision scheme on type-1 Sommerville tetrahedra. All of these traversals leave something to wish for.

Maehara's subdivision scheme generalizes the recursive tessellation that underlies P\'olya's space-filling curve for isosceles right trangles. However, in higher dimensions, the Maehara-reflected traversal proposed in this paper does not maintain continuity. Can we find an elegant generalization of the P\'olya curve (possibly with a different underlying tessellation) that is continuous in any number of dimensions?

\begin{wrapfigure}{r}{0.25\textwidth}
  \vspace{-18pt}
  \includegraphics[width=\hsize,page=2]{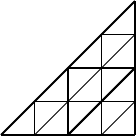}
  \vspace{-15pt}
  \caption{Two levels of the underlying tiling of Hill-Z in 2D.}
  \label{fig:HillZ2D}
  \vspace{-24pt}
\end{wrapfigure}
With the Hill-Z traversal, everything seemed to fall into place in a natural way, but does the curve really need to jump around as much as it appears to do in the drawing in Figure~\ref{fig:hillztraversal}? Could we construct a more elegant traversal while still keeping to Freudenthal's subdivision scheme? Note that already in two dimensions, this tiling (see Figure~\ref{fig:HillZ2D}) is challenging. A symmetric or a face-continuous traversal based on this tiling is not possible, and finding a semi-face-continuous traversal (or proving that none exist) might be a non-trivial task. Burstedde et al.~\cite{BHI17} show that the two-dimensional traversal $\descr{\fwd[1\\2]\edge[1]\fwd[1\\2]\rev[\m1\\\m2]\edge[2]\fwd[1\\2]}$ from Burstedde and Holke~\cite{BuH16} is not semi-face-continuous.

Regarding the type-1 Sommerville tetrahedra, here, too, the investigation should perhaps start in two dimensions. The Sommerville tetrahedron can be described as a special type of Hill simplex that I would call a \emph{cycloscheme}. It can be defined by a set of vectors $v_0,...,v_d$ of equal length, such that, for any $i \neq j$, the angle between $v_i$ and $v_j$ is $\arccos(-1/d)$, and moreover, $\sum_{i=0}^d v_i = 0$. The vertices $p_0,...,p_d$ of the cycloscheme (modulo translation) are now given by $p_h = p_0 + \sum_{i=1}^h v_i$ (see also Maehara~\cite{Mae15}). The two-dimensional cycloscheme is the equilateral triangle. What traversals with favourable properties can we construct for equilateral triangles? What adaptations do we need to make to our notation method to deal with traversals of equilateral triangles? In three dimensions, the symmetries of the cycloscheme are a subset of the symmetries of an appropriately aligned cube, but in two dimensions they are not. We could consider using signed permutations in combination with barycentric coordinates. Would that also be necessary and sufficient to encode the required transformations for cycloschemes in four and more dimensions?

%\HHX{
%rectangle-filling curves (1:sqrt 2, 1:sqrt 3, 2:3)
%}

\paragraph{Zooming out}
In some applications, we may want to order points of which the range of possible coordinate values is not known beforehand. This makes it problematic to scale the data to lie within the unit cube. It would therefore be useful to extend the traversals from the unit cube or simplex to all of $d$-dimensional space. Clearly this is possible for all recursively defined traversals: we can always zoom out by regarding the unit cube or simplex as a part, say the $i$-th part, of a larger unit whose traversal is rotated, reflected and/or reversed such that it is consistent with the definition of the traversal of the original unit. However, implementing this approach requires care: the result depends on what value we choose for $i$. In fact, for some curves, all of $d$-dimensional space can only be filled if we do not take the same value for $i$ each time we zoom out. It might be good to decide on a standard way to zoom out for each curve, after looking more carefully into the effect that different solutions may have on practical implementations. In any case, for the traversals in Section~\ref{sec:discontinuous}, we would want to ensure that it does not become more complicated to implement them by ``bit tricks'' on the coordinate matrix.

If we want to be able to compute an inverse of the traversal (a function that maps any point in $d$-dimensional space to a position $t$ in the traversal), tie-breaking needs special intention. Zooming out can have the result that some points on the boundary of the original unit cube or simplex are now treated as part of an adjacent cube or simplex, and it may be problematic if the position of these points along the traversal changes as a result.

A more fundamental issue arises with the Harmonious Hilbert curve and the curves presented in Section~\ref{sec:Peano}. The curves as presented in this article have the following property: the points of the unit cube that lie on any $(d-1)$-dimensional facet touching the lexicographically smallest corner of the cube, are visited in the order of the corresponding $(d-1)$-dimensional curve. If we zoom out, we might want this property to extend to the full $(d-1)$-dimensional axis-parallel hyperplanes through the lexicographically smallest corner of the cube, but it is not clear whether this can be realized with each of these curves~\cite{Hav12}.

\subsection*{Acknowledgements}
I thank Arie Bos who originated the notation system without which this article could not have been written; Carsten Burstedde for sending me his proof of the semi-face-continuity of the Z-traversal; Michael Bader and Tobias Weinzierl for posing the problems that ultimately led to the Inside-out traversal and the palindromic tetrahedral traversal; Zuzana Pat\'akov\'a for pointing me to Maehara's work and other useful sources; Freek van Walderveen for discovering the first algorithm to generate Harmonious Hilbert curves in any number of dimensions; and Julian Rohrhuber for our inspiring discussions which prompted me to write this all up.

\break
\bibliographystyle{abbrv}

\clearpage
\appendix

\section{The squared Meander curve is not self-similar}\label{apx:notselfsimilar}
\small
Consider the Meander curve, translated to fill the square $[0,1]^2$:

\addvspace\baselineskip
\hbox to\hsize{\hfill\includegraphics{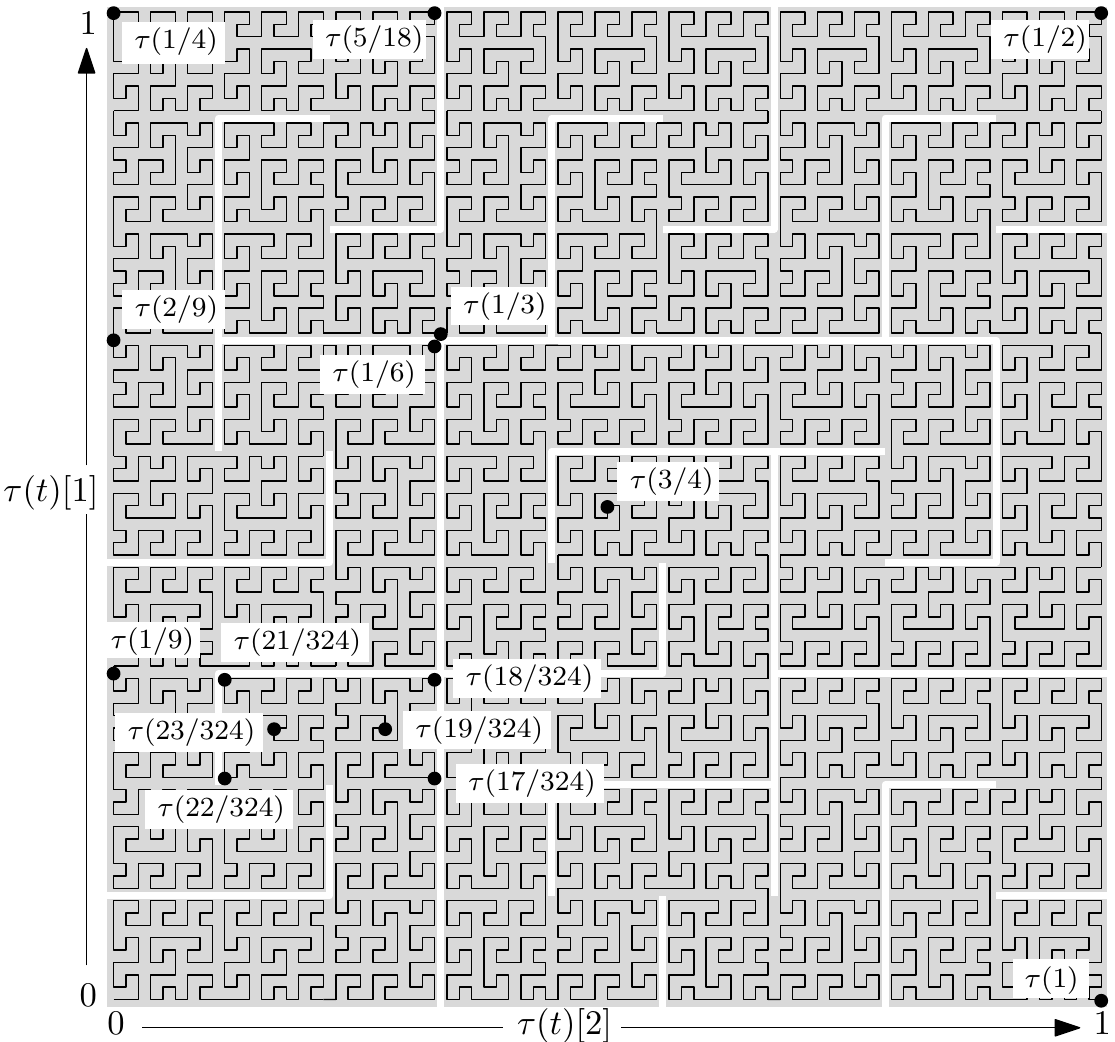}\hfill}
\addvspace\baselineskip

Consider the second level of subdivision, that is, the subdivision into 81 squares.
One quarter, two quarters and three quarters into the fifth subsquare, with transformation $[2,1\}$, we find the points:\[
\tau\left(\frac{17}{324}\right) = \left(\frac29,\frac13\right);\quad \tau\left(\frac{18}{324}\right) = \left(\frac13,\frac13\right);\quad \tau\left(\frac{19}{324}\right) = \left(\frac5{18},\frac5{18}\right).\]
Applying $\tau$ to each coordinate, we find:\[
\tau'\left(\frac{17}{324}\right) = \left(\frac23,0,\frac23,\frac13\right);\quad \tau'\left(\frac{18}{324}\right) = \left(\frac23,\frac13,\frac23,\frac13\right);\quad \tau'\left(\frac{19}{324}\right) = \left(1,\frac13,1,\frac13\right).\]
Note that the point in the middle, $\tau'(18/324)$, shares at least two coordinates with each of the other two points.
Now consider the points at one quarter, two quarters and three quarters into the sixth of 81 subsquares, with transformation $[-2-1\}$, which induces two top-level reflections. We have:\[
\tau\left(\frac{21}{324}\right) = \left(\frac13,\frac19\right);\quad \tau\left(\frac{22}{324}\right) = \left(\frac29,\frac19\right);\quad \tau\left(\frac{23}{324}\right) = \left(\frac5{18},\frac16\right).\]
Applying $\tau$ to each coordinate, we find:\[
\tau'\left(\frac{21}{324}\right) = \left(\frac23,\frac13,\frac13,0\right);\quad \tau'\left(\frac{22}{324}\right) = \left(\frac23,0,\frac13,0\right);\quad \tau'\left(\frac{23}{324}\right) = \left(1,\frac13,\frac23,\frac13\right).\] The last two points do not share any coordinates, so the traversal in the sixth of the 81 subcubes of the 4-dimensional unit cube cannot be similar to the traversal in the fifth of the 81 subcubes of the 4-dimensional unit cube---not even under reversal.

\end{document}